\documentclass[nofigure]{pasj00}
\usepackage{graphicx}
\tenpoint

\SetRunningHead{A Dynamical Model of the Origin of the 
Local Group of Galaxies}{T. Sawa and M. Fujimoto}

\title{A Dynamical Model for the Orbit \\
of the Andromeda Galaxy M31\\
and the Origin of the Local Group of Galaxies}
\author{Takeyasu \textsc{Sawa}}
\affil{%
   Department of Physics and Astronomy,
   Aichi University of Education, Kariya 448-8542}
\email{tsawa@auecc.aichi-edu.ac.jp}
\and
\author{Mitsuaki \textsc{Fujimoto}}
\affil{%
   Temporary address: U-Lab, Department of Physics and Astrophysics, 
Nagoya University, Nagoya 464-8602}
\email{fujimoto@a.phys.nagoya-u.ac.jp}
\KeyWords{galaxies: formation --- 
galaxies: kinematics and dynamics --- 
galaxies: Local Group}
\Received{$\langle$reception date$\rangle$}
\Accepted{$\langle$acception date$\rangle$}
\Published{$\langle$publication date$\rangle$}
\begin{document}
\maketitle

\begin{abstract}
We propose a new model for the origin and evolution of the Local
Group of Galaxies (LGG) which naturally explains the formation
of the Magellanic Clouds and their large orbital angular momenta
around the Galaxy. The basic idea is that an off-center hydrodynamical
collision occurred some 10 Gyr ago between the primordial gas-rich 
Andromeda galaxy and the similar Galaxy, and compressed the halo gas to
form the LGG dwarf galaxies including the Magellanic Clouds. In this
model, new-born dwarf galaxies can be expected to locate
near the orbital plane of these two massive galaxies.
In order to see the reality of this model, we
reexamine the two-dimensional sky distribution of the LGG 
members and the Magellanic Stream, we confirm an earlier and 
widely-discussed idea that they align along two similar great 
circles, each with an angular width of 
$\sim $30$^{\circ}$, and the planes of these circles
are approximately normal to the line joining the present 
position of the sun and the Galactic center. Further we make a
three-dimensional distribution map of these objects, 
and observe it from various directions. A well-defined 
plane of finite thickness is found, within which most of 
the member galaxies are confined, supporting the existence 
of the above circles on the sky. 
Thus we could determine the orbital elements of
M31 relative to the Galaxy through reproducing the 
well-studied dynamics of the LMC and the SMC around 
the Galaxy. The expected proper motion of M31 is 
$(\mu _{l},\ \mu_{b})=(38\ \mu{\rm as~yr}^{-1}, 
-49\ \mu{\rm as~yr}^{-1})$.
Probable orbital motions of the other dwarfs are
also determined, and the
corresponding proper motion for each object is given 
to compare with observations in near future.
\end{abstract}

%
\section{Introduction}

In spite of extensive observational/theoretical studies
of the Magellanic Clouds, their origins remain
unsolved for a long time. Did they initially form as satellite
galaxies of the Galaxy, or did they fall into the Galaxy in
the course of dynamical evolution of the LGG \citep{byr94}.
 A more realistic hint to this problem can be
presented by studies of the origin of the Magellanic Stream
(MS). Tidal models have been successfully introduced to the
dynamics of the Galaxy-LMC-SMC system for reproducing the
geometrical and dynamical structure of the MS 
\citep{mur80,lin82,gar94,gar96}.
In such models, orbits of the Large Magellanic Cloud (LMC) 
and the Small Magellanic Cloud (SMC) can be traced back in time 
over the entire past period of $\sim 10$ Gyr: 
The orbital plane is approximately perpendicular 
to the line joining the present 
position of the sun and the Galactic center,
and they are viewed to move counterclockwise along a nearly 
great circle centered on 
$(l, b)=(0^{\circ},0^{\circ})$ or the Galactic center.

However, the tidal model in which the Magellanic Clouds formed 
in the neighborhood of the Galaxy cannot explain large
orbital angular momenta of the Magellanic Clouds around the
Galaxy (see, for example, \citet{fic94}). 
According to the theory of the Magellanic
Stream, the apo-Galactic distance of the LMC orbit is 
$R_\mathrm{LMC} \sim 200$ kpc and 
its transverse velocity is $V_\mathrm{LMC} \sim $100 km s$^{-1}$ some 10 Gyr 
ago, implying that if the LMC mass is 
$M_\mathrm{LMC} \sim 2 \times 10^{10}\MO$, the
orbital angular momentum of the Magellanic Clouds $L_\mathrm{LMC}$ at that time  
is given by 
\begin{eqnarray}
L_\mathrm{LMC}
& \sim & M_\mathrm{LMC}R_\mathrm{LMC}V_\mathrm{LMC} \nonumber \\
& \sim & 2 \times 10^{10}\MO \times 200{\rm ~kpc~}\times 100 {\rm ~km~s}^{-1} \nonumber \\
& = & 4\times 10^{14} \MO {\rm ~kpc~km~s}^{-1},
\end{eqnarray}
while the spin of the Galactic disk $L_\mathrm{disk}$ 
in flat rotation with velocity $V_{0}$ is given by
\begin{eqnarray}
L_\mathrm{disk}
& \sim & \int r V_{0} dm \nonumber \\
& \sim & \int_{0}^{R_\mathrm{disk}}
\int_{0}^{2\pi}\int_{-h/2}^{h/2} \rho_{0}r^{2}V_{0}drd\phi dz \nonumber \\
& = & \frac{2\pi}{3}\rho_{0}V_{0}R^{3}h,
\end{eqnarray}
where $\rho_{0}$ is the density of the disk, $h$ the thickness, 
and $R_\mathrm{disk}$ the outer radius.
In this estimate, we assume $\rho_{0}$ and $h$ to
be constant.
Since the total mass of the disk $M_\mathrm{disk}$ is given by 
$M_\mathrm{disk}= \pi R_\mathrm{disk}^{2} h \rho_{0}$, 
the spin of the Galactic disk $L_\mathrm{disk}$ is estimated as
\begin{eqnarray}
L_\mathrm{disk}
& \sim & \frac{2}{3}M_\mathrm{disk}R_\mathrm{disk}V_{0} \nonumber \\
& \sim & \frac{2}{3}\times2\times 10^{11} \MO 
\times 15 {\rm kpc} \times 220{\rm ~km~s}^{-1} \nonumber \\
& = & 4\times 10^{14} \MO{\rm ~kpc~km~s}^{-1},
\end{eqnarray}
if we adopt $V_{0}=220$ km s$^{-1}$ and $M_\mathrm{disk}=2 \times 10^{11} \MO$ for
$R_\mathrm{disk}=15$ kpc. We find that the
orbital angular momentum of the LMC, $L_\mathrm{LMC}$, is 
comparable to the spin of the Galactic disk $L_\mathrm{disk}$.
It should be also born in mind that these 
angular momenta make a right angle against 
each other. They cannot be explained in terms of the
tidal interaction in the LGG \citep{got78}.
Further it has also been widely recognized that the morphological 
type of dwarf galaxies near their massive parent galaxy are
generally dwarf spherical or spheroidal, but that
the Magellanic Clouds are morphologically classified as dwarf irregulars
in spite of their close location to the Galaxy and are
exceptionally massive \citep{ein74} and gas-rich. 

These facts are difficult to understand only 
in terms of the
gravitational interaction among the three-body system, 
the Galaxy-LMC-SMC. Then \citet{fuj99} and 
\citet{saw99} conducted numerical
simulations of the Magellanic Stream by taking into account the
gravitational effect of the nearby massive galaxy M31 which 
is approaching us with velocity of 
120 km s$^{-1}$ if we adopt the standard rotation velocity of the LSR
($V_{0}=220$ km s$^{-1}$) and the solar motion 
($v_{\odot}=16.5$ km s$^{-1}$, 
$l_{\odot}=53^{\circ}$, $b_{\odot}=25^{\circ}$). 
With the orbital elements of M31 
assumed by \citet{kah59}, \citet{fuj99}, and \citet{saw99}, 
it is found that the binary orbits of the LMC 
and the SMC at their early times are much disturbed compared with
those determined by \citet{mur80} 
and \citet{gar94}, but no successful result has been
obtained for solving the 
angular momentum problem even through the 
four-body dynamics of the Galaxy-M31-LMC-SMC.

The present authors come to take a view point that 
some questions about the Galaxy-LMC-SMC system must be answered through 
a more-global model for the 
LGG in which
the Galaxy, the LMC and the SMC should be treated as three 
objects among more-than-forty members of the LGG.
For this reason we begin with in the next section our 
reexamination of the sky (two-dimensional) and spatial
(three-dimensional) distributions of the member galaxies 
of the LGG, which would give key ideas for the 
dynamics and the origin of the LGG.

In concluding this section, we note that
many studies have tried to determine the past orbits
of the LGG members under their various assumptions.
They are conducted mostly in virial scheme and, therefore,
the results are rather stochastic. Here we should 
refer to three important works
by \citet{mis85}, \citet{pee93}, and \citet{lyn99}.
They have pioneered the global LGG dynamics, although
many basic differences are present from our ideas 
and results, particularly concerning the origin of the LGG.

\section{Characteristic Distributions of the Local Group Dwarf 
         Galaxies}

\subsection{Distributions of the LGG Members in the Galactic 
   Coordinates}     

The concept of a Local Group of Galaxies was introduced 
as early as at the time of Hubble's distance and
redshift measurements of extragalactic objects.
The number of member galaxies of the Local Group has increased 
since that time and it is still so even at present. 
It is more than forty: See \citet{mat98}
and \citet{van00} for reviews of 
the LGG. 

\begin{figure*}
 \includegraphics[width=17cm]{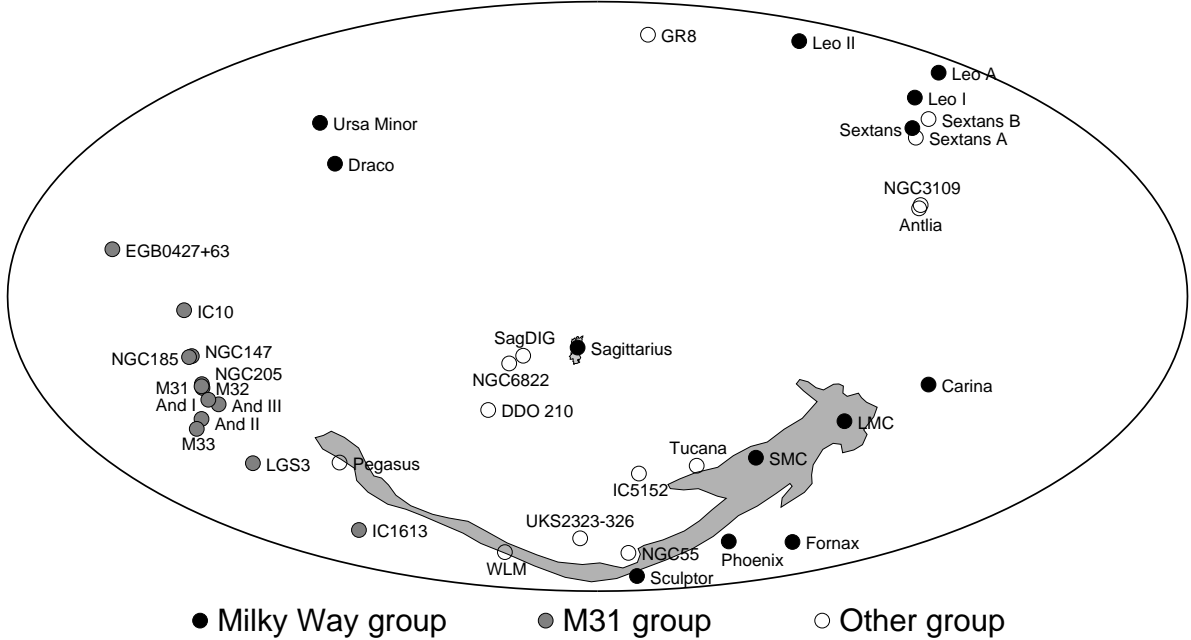}
\caption{Two-dimensional distribution of the members 
of the LGG and the Magellanic
Stream in Hammer's projection onto the sky. 
The data are due to
\citet{mat98}, and the lines of the Galactic longitudes
and latitudes are not given here 
in order to emphasize that these objects 
are distributed not in random but in some systematic way 
like in a ring of large diameter, except four dwarfs 
in the direction of the Galactic center. The filled
circles indicate the Galaxy group members, the netted 
circles the M31 group and the open circles the 
nonparent group. Note that a tidal tail is sketched
at the Sagittarius dwarf. Although it
is very small, its major axis makes evidently a 
nearly right angle against the galactic plane. }
\label{fig-1}
\end{figure*}

\begin{figure*}
 \includegraphics[width=17cm]{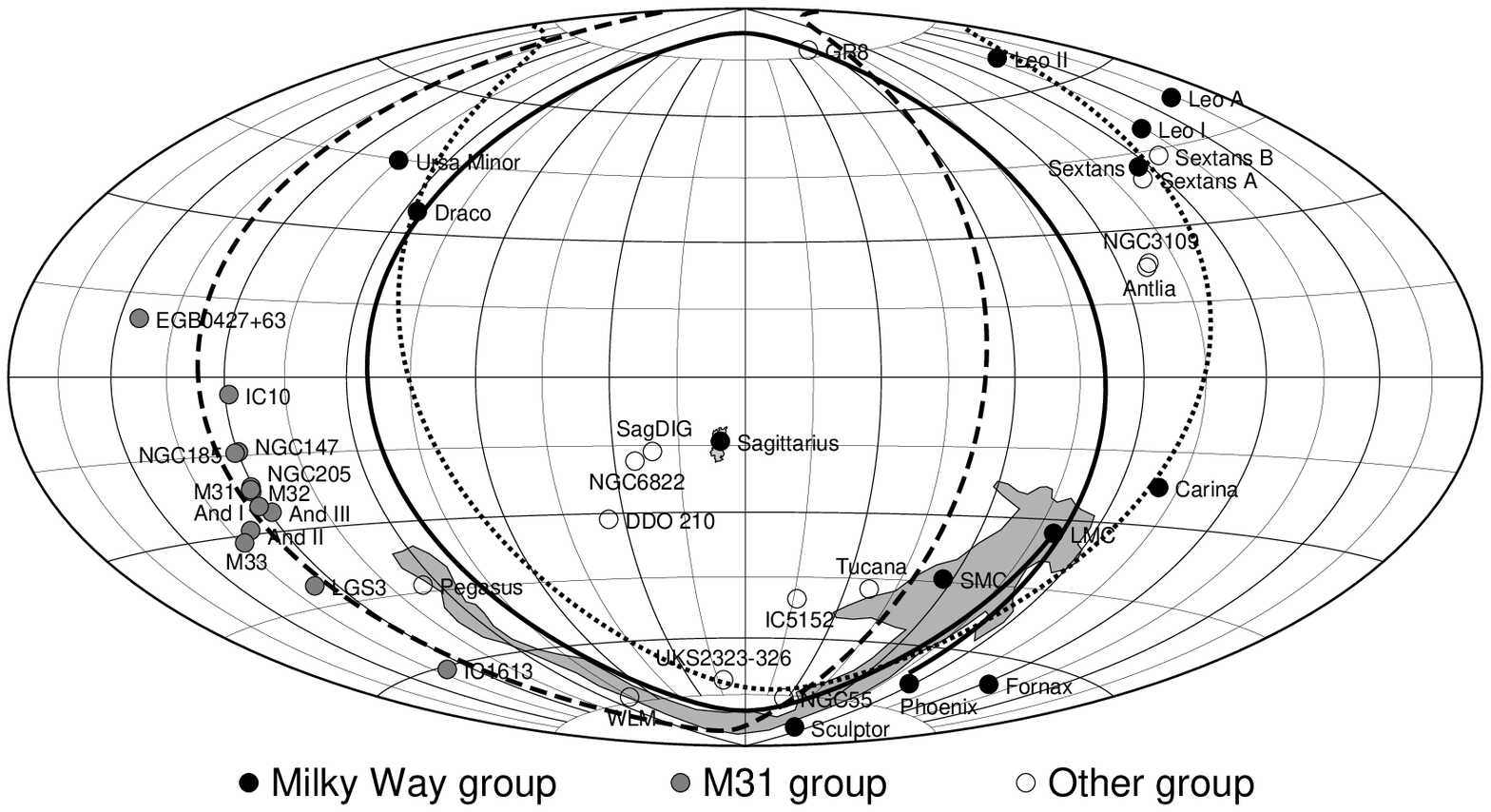}
\caption{Two great circles in dashed and dotted lines and another
circle in solid line on the $(l, b)$-plane of the sky.  
The constant lines of the Galactic longitude and latitude 
$(l, b)$ cover all the same sky as in figure \ref{fig-1}.
The dotted line indicates a great 
circle fitted to the distribution of the Galaxy group 
members, the dashed line to that of the M31 group members 
and the solid line traces a last revolution orbit of the LMC
about the Galaxy \citep{gar94}.
These three circles occupy approximately the same 
large area of the sky, suggesting the existence of the
ring-like sky distribution of the LGG members, although 
the angle discordance of $\sim 30^{\circ}$ in maximum
is recognized among these circles.}
\label{fig-2}
\end{figure*}

Figure \ref{fig-1} shows 
a sky distribution of the LGG members listed 
in the up-to-date table of \citet{mat98}.
The galactic coordinates ($l,b$) is adopted,
but longitude/latitude grid-lines
are ignored here in order 
to emphasize an impression that the LGG members are  
distributed not at random but in a somewhat systematic way 
forming a ring of large diameter of about 180$^{\circ}$.
The major axis of the sky rimmed with an ellipse
is the Galactic plane, with $(l, b)=(180^{\circ},0^{\circ})$
at the both extreme ends, and the minor axis links the north 
and south Galactic poles. The Galactic center 
$(l, \ b)=(0^{\circ},0^{\circ})$ is located at the
geometrical center of this ellipse.

If we exclude four
dwarfs in the direction of the Galactic center, 
Sagittarius, SagDIG, NGC6822 and DDO210, 
the existence of the ring of $\sim 30^{\circ}$ in width
appears to be more evident. The plane of the ring 
is perpendicular both to the Galactic plane and 
to the line joining the present position of the sun
and the Galactic center. The apparently small tidal
structure of the Sagittarius will be discussed later.

Such a ring-like distribution of the LGG dwarf 
galaxies and the Magellanic Stream
was already pointed out in a few previous studies until the
end of the 1970s, but it is widely considered to be
rather accidental with no special meanings except
some studies. As will be touched upon later, 
\citet{kun79} claimed that the 
ring-like distribution of the Galaxy group dwarfs is 
an evidence for a big dynamical event which occurred in 
the neighborhood of the Galaxy some billion years 
ago and that a few dozen of 
proto-dwarf gaseous debris were driven to form and scatter 
on the plane as observed today 
in figure \ref{fig-1}.  

\begin{table*}[t]
\caption{Applied Parameters for the Local Group of Galaxies$^{*}$.
\label{tab-1}}
{\footnotesize 
\begin{center}
\begin{tabular}{llrrrrl} \hline \hline
Galaxy Name     & Other name    &  $l$~~ & $b$~~    & $r$~~&$v_{r}$~~&  Group  \\ \hline
WLM 		& DDO 221 	&  75.9  & $-73.6  $& 925  &$ -123 $& LGC      \\
NGC55 		&   		&  332.7 & $-75.7  $& 1480 &$ 124  $& LGC      \\
IC10  		& UGC192 	&  119.0 & $-3.3   $& 825  &$ -342 $& M31      \\
NGC147 		& DDO3 		&  119.8 & $ -14.3 $& 725  &$ -193 $& M31      \\
And III 	&  		&  119.3 & $ -26.2 $& 760  &  ---~  & M31      \\
NGC185 		& UGC396 	&  120.8 & $ -14.5 $& 620  &$ -204 $& M31      \\
NGC205 		& M110 		&  120.7 & $ -21.1 $& 815  &$ -229 $& M31      \\
M32 		& NGC221 	&  121.2 & $ -22.0 $& 805  &$ -197 $& M31      \\
M31 		& NGC224 	&  121.2 & $ -21.6 $& 770  &$ -297 $& M31      \\
And I 		&    		&  121.7 & $ -24.9 $& 805  &  ---~  & M31      \\
SMC  		& NGC292 	&  302.8 & $ -44.3 $& 58   &$ 148  $& MW       \\
Sculptor	&  		&  287.5 & $ -83.2 $& 79   &$ 102  $& MW       \\
LGS3 		& Pisces 	&  126.8 & $ -40.9 $& 810  &$ -272 $& M31      \\
IC1613 		& DDO8 		&  129.8 & $ -60.6 $& 700  &$ -234 $& M31/LGC  \\
And II 		&    		&  128.9 & $ -29.2 $& 525  &   ---~ & M31      \\
M33 		& NGC598 	&  133.6 & $ -31.3 $& 840  &$ -181 $& M31      \\
Phoenix 	&   		&  272.2 & $ -68.9 $& 445  &$ 56   $& MW/LGC   \\
Fornax 		&   		&  237.1 & $ -65.7 $& 138  &$ 53   $& MW       \\
EGB0427+63 	& UGCA92 	&  144.7 & $ +10.5 $& 1300 &$ -87  $& M31      \\
LMC 		&   		&  280.5 & $ -32.9 $& 49   &$ 274  $& MW       \\
Carina 		&   		&  260.1 & $ -22.2 $& 101  &$ 224  $& MW       \\
Leo A 		& DDO69 	&  196.9 & $ +52.4 $& 690  &$ 26   $& MW/N3109 \\
Sextans B 	& DDO70 	&  233.2 & $ +43.8 $& 1345 &$ 303  $& N3109    \\
NGC3109 	& DDO236 	&  262.1 & $ +23.1 $& 1250 &$ 404  $& N3109    \\
Antlia 		&    		&  263.1 & $ +22.3 $& 1235 &$ 361  $& N3109    \\
Leo I 		&    		&  226.0 & $ +49.1 $& 250  &$ 286  $& MW       \\
Sextans A 	& DDO75 	&  246.2 & $ +39.9 $& 1440 &$ 325  $& N3109    \\
Sextans 	&   		&  243.5 & $ +42.3 $& 86   &$ 277  $& MW       \\
Leo II 		& DDO93 	&  220.2 & $ +67.2 $& 205  &$ 76   $& MW       \\
GR8 		& DDO155 	&  310.7 & $ +77.0 $& 1590 &$ 215  $& GR8      \\
Ursa Minor 	& DDO199 	&  105.0 & $ +44.8 $& 66   &$ -248 $& MW       \\
Draco 		& DDO208 	&  86.4  & $ +34.7 $& 82   &$ -293 $& MW       \\
Milky Way 	&  		&  0.0   & $ 0.0   $& 8.5  &  ---~  & MW       \\
Sagittarius 	&   		&  5.6   & $ -14.1 $& 24   &$ 140  $& MW       \\
SagDIG 	 	& UKS1927-177 	&  21.1  & $ -16.3 $& 1060 &$  -79 $& LGC      \\
NGC6822 	& DDO209 	&  25.3  & $ -18.4 $& 490  &$ -54  $& LGC      \\
DDO 210 	& Aquanus 	&  34.0  & $ -31.3 $& 800  &$ -137 $& LGC      \\
IC5152 		&  		&  343.9 & $ -50.2 $& 1590 &$ 124  $& LGC      \\
Tucana 		&  		&  322.9 & $ -47.4 $& 880  &  ---~  & LGC      \\
UKS2323-326 	& UGCA 438 	&  11.9  & $ -70.9 $& 1320 &$ 62   $& LGC      \\
Pegasus 	& DDO216 	&  94.8  & $ -43.5 $& 955  &$ -182 $& LGC      \\ \hline \hline
\end{tabular}
\end{center}
$*)$ Due principally to \cite{mat98}, but the radial velocities
of the LMC and the SMC are to \cite{tul88}. 

Galaxy Name: Names of the Members of the Local Group of Galaxies. 
Other name: Other names of the applied members. 

$l$: The Galactic longitude in degree. 

$b$: The Galactic latitude in degree. 

$r$: Distance from the sun measured in kpc. 

$v_{r}$: Heliocentric radial velocity in km s$^{-1}$. 

Group: Name of galaxy group, to which the above members belong. MW means 
the Milky Way, M31 the Andromeda galaxy.

}
\end{table*}

In order to emphasize the non-random distribution of
dwarfs found in figure \ref{fig-1} more quantitatively, 
we draw in figure \ref{fig-2} two great circles and another
great-circle-like line on the $(l, b)$ plane.
The dotted line is a great circle defined by twelve dwarf galaxies 
of the Galaxy group except Sagittarius which locates near the
Galactic center, while the dashed line is that 
defined by the M31 group, i.e. M31 and nine dwarfs.
The great-circle-like solid line represents the last 
previous ($\sim 2$ Gyr) orbit of the LMC projected onto
the sky, which traces the Magellanic 
Stream. For the data of the LGG members, see Table \ref{tab-1}.

These two great circles and another great-circle-like
line (the LMC's orbit) do not coincide with 
each other exactly, but they occupy roughly a same half 
area of the sky. The existence of the ring-like 
structure suggested in figure \ref{fig-1} is 
considered as a result of superposition of these 
circles in figure \ref{fig-2}.

The discordance among the three circles in 
figure \ref{fig-2}, and the four dwarfs in 
the direction of the Galactic center 
in figure \ref{fig-1} will be taken into consideration in
the framework of our LGG model presented later.

\subsection{Three-Dimensional Space Distributions of the 
Local Group of Galaxies}     

Figures \ref{fig-1} and \ref{fig-2} give only
distribution of the LGG dwarfs projected onto the sky
planes, lacking the information on 
the distance measured on the
line-of-sight. We plot the positions of the LGG members 
(Table \ref{tab-1}) in the
three-dimensional rectangular coordinates and see 
them from various directions. Such analyses have 
been made in many other previous studies, and are known to be
convenient to obtain an overview about the LGG 
structure because we can 
avoid the large parallaxes of the members which are 
located close to the Galaxy. The dwarf galaxies 
in the Sagittarius region, which are discarded in our 
discussions about the ring structure in figures \ref{fig-1} 
and \ref{fig-2}, are this case. See \citet{maj94}, \citet{mat98},
\citet{lyn99}, and \citet{van00} for
their review articles about the three-dimensional structure
of the LGG.

\begin{figure*}
 \includegraphics[width=17cm]{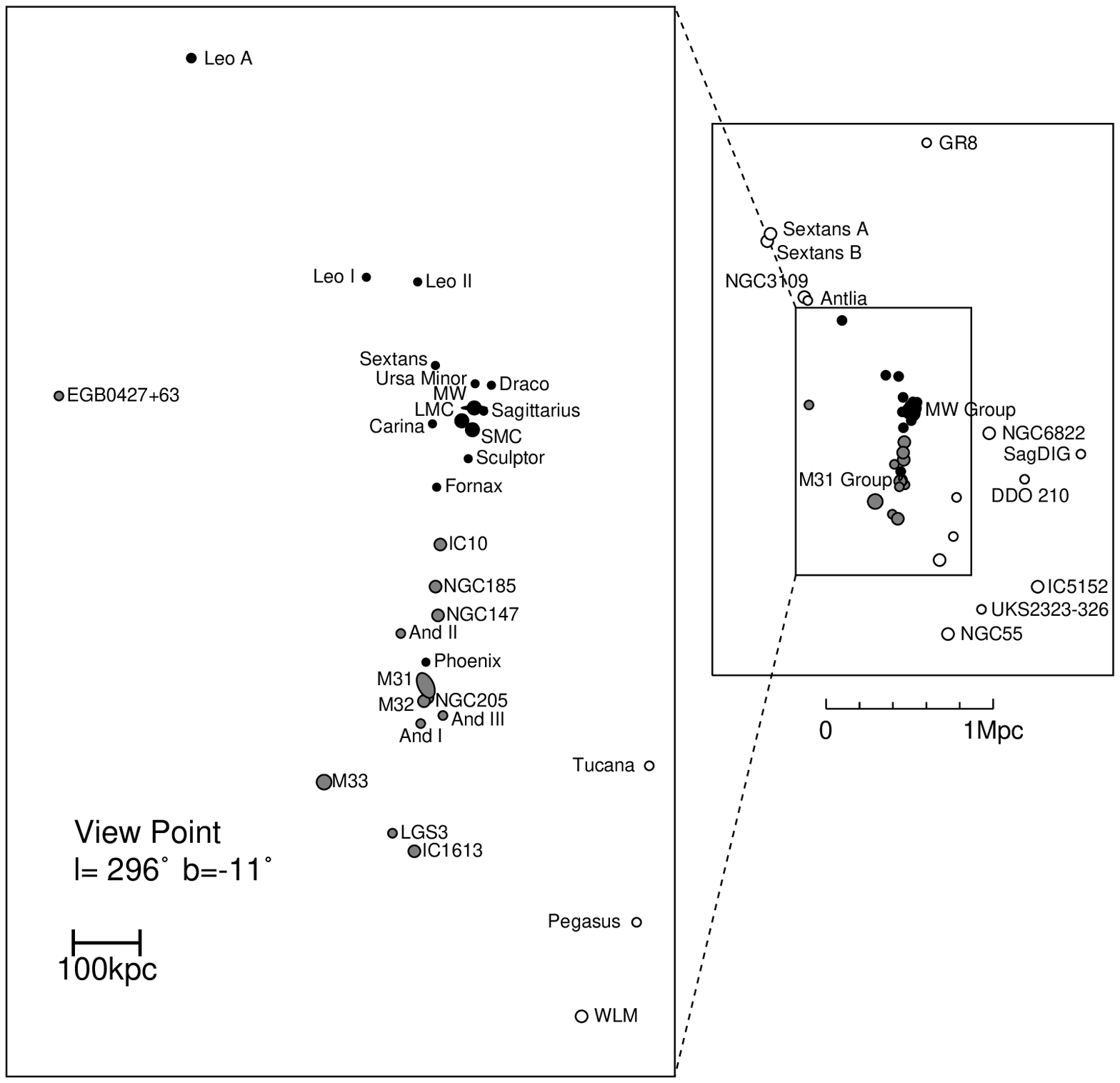}
\caption{Three-dimensional space distribution of the LGG
members, seen from the direction 
$(l, b)=(296^{\circ},-11^{\circ})$. This is an 
{\it edge-on view} of the orbital plane of the Galaxy 
and M31. The perspective 
representation is not applied.
The filled circles denote the Galaxy group members, the 
netted ones the M31 group members and the open ones the members
that belong neither to the Galaxy nor to M31. We recognize 
that the LGG members are distributed in a flat disk 
of finite thickness of 50-100 kpc.
The size of each circle represents its brightness only 
qualitatively. Note that some number of dwarf galaxies
are not given on the outside of the LGG region
whose radius is $\sim $1Mpc. They are plotted
in the inset on upper-right, together with
all galaxies listed in the Table 1.}
\label{fig-3}
\end{figure*}

Figure \ref{fig-3} shows a distribution of the LGG members seen from the 
direction $(l, b)=(296^{\circ}, -11^{\circ})$. There can be seen clear
alignment along the line passing M31 and the Galaxy;
the LGG members are distributed in a coplanar way
or in a flat disk of finite thickness of 50-100 kpc, which
makes the ring-like or circle-like distributions of the 
LGG members in figures \ref{fig-1} 
and \ref{fig-2}. The disk plane is approximately 
perpendicular to 
the direction toward $(l, b)= (0^{\circ}, 0^{\circ})$,
and/or the Galactic center. 

Three objects shown by open circle locate 200-300 kpc apart from
the disk plane. Compared with the radius of the LGG
of 1Mpc \citep{lyn99}, such separation is not so large,
and hence we think that they also contribute to defining the LGG plane. Other 
two objects, EGB0427+63 and Leo A, are exceptionally distant
from the disk plane; they may be intruders (see later).

\citet{har00} analyzed the 3-dimensional positions
of the LGG members and determined the spheroidal
distribution for the Galaxy group members and the M31 group ones,
respectively. He pointed out that the disk plane of these two
spheroids are nearly parallel, but make a
small angle between them. Indeed, if we observe figure \ref{fig-3} 
from a slightly different view-direction,
the Galaxy members align more clearly and those of the 
M31 group are a little more dispersed.

Although such structure difference may indicate some special
initial condition for the LGG dynamics \citep{har00},
we ignore it here and construct a dynamical model for
the LGG. Hartwick's fine structure will be 
considered in another paper as a
first-order perturbation to our LGG model. Then,
Raychaudhury and Lynden-Bell's dynamics \citep{ray89} may
give us a new hint to this problem that the tidal torque
due to nearby massive galaxies 
outside the LGG region would have 
perturbed the motions of M31, the Galaxy and their dwarfs over the  
past $10^{10}$ years.

\begin{figure*}
 \includegraphics[width=17cm]{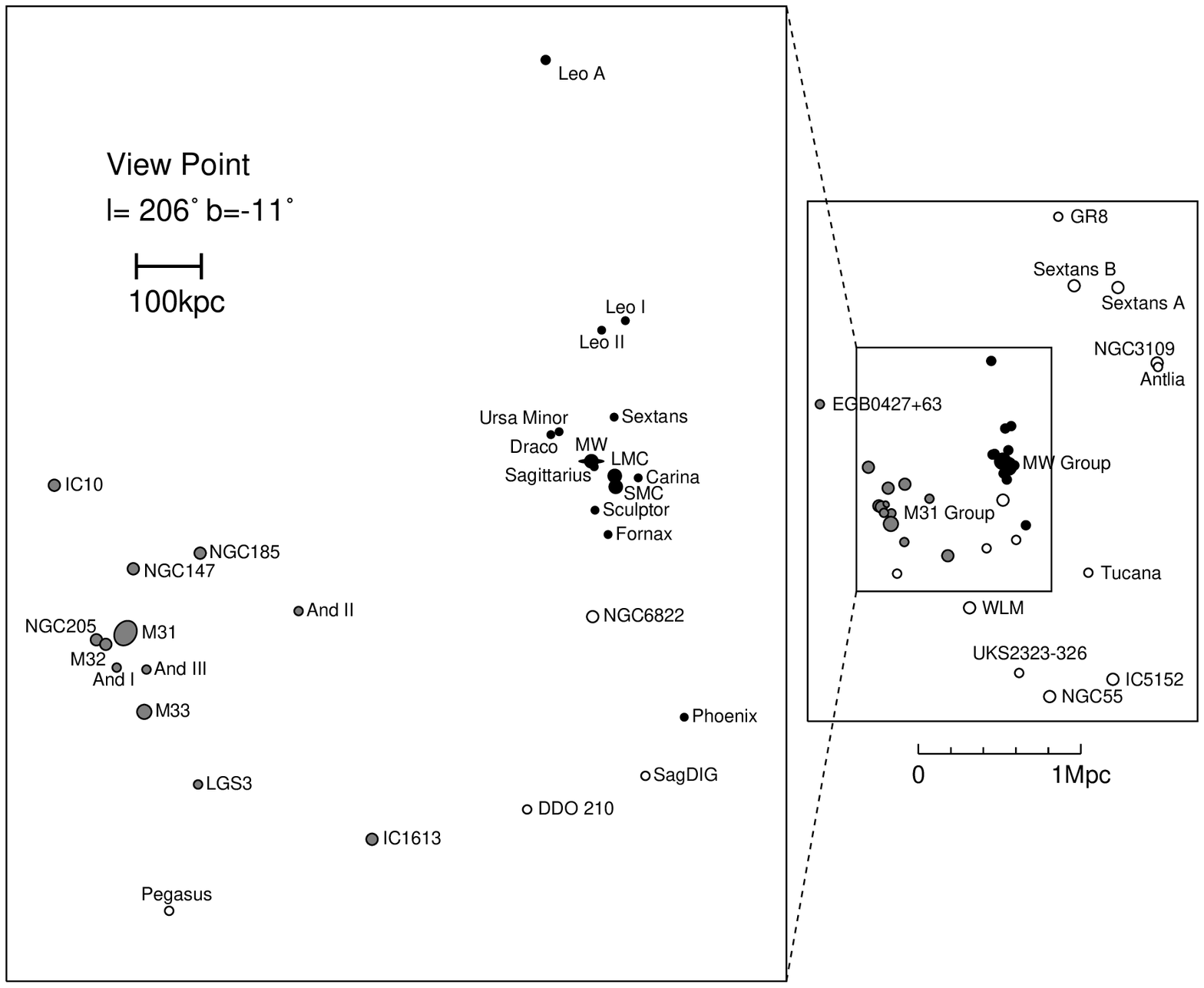}
\caption{The same as in figure \ref{fig-3}, but seen from
the direction $(l, b)=(206^{\circ},-11^{\circ})$. This is
a {\it face-on view} of the LGG plane, or of the orbital 
plane of the Galaxy and M31 in the present model.
The Galaxy group and the M31 group members, and nongroup
members appear to form a fragmentary ring whose 
diameter is $\sim 800$kpc and more. 
We note that some of the nongroup members, Pegasus, 
DDO210, SagDIG and NGC6822 are not exactly
in the LGG plane of finite thickness. They are located 
off from the plane by 160-730 kpc, large compared with
the disk thickness, but small compared with the LGG's
radius of $\sim 1$ Mpc. They also contribute to forming our 
model of the coplanar structure of the 
LGG. The inset on upper-right is the same as
that in figure 3, but viewed face-on.}
\label{fig-4}
\end{figure*}

Figure \ref{fig-4} represents a 3-dimensional distribution of the LGG members
viewed from the direction $(l, b)$
=(206$^{\circ},\ -11^{\circ}$), normal to the 
plane identified in figure \ref{fig-3}.
We call it a 
{\it face-on view} against the plane of the LGG, and then 
figure \ref{fig-3}, an {\it edge-on view} against the LGG disk.
Even in the face-on view, the LGG members are
distributed not in random but in a systematic way, as
grouped around M31 and the Galaxy, respectively.

Dwarf galaxies of open circles gather around neither M31
nor the Galaxy. They appears to link these 
two mass-dominant galaxies, suggesting that M31 and the 
Galaxy interacted strongly at their early ages. However,
we note that these dwarfs are not confined exactly in the LGG disk
(the plane of this page) of finite thickness but are apart
from it with some distance.

\section{A Dynamical Model for the Local Group 
         of Galaxies}

The ring-like or circular distributions of the LGG 
members on the sky (figures \ref{fig-1} and \ref{fig-2}) are
consistent with the existence of the LGG disk 
with finite thickness of 50-100 kpc 
(figures \ref{fig-3} and \ref{fig-4}).

As briefly touched upon in section 2, 
\citet{kun79} considered that a destructive
dynamics occurred some billion
years ago in the space near the Galaxy and
a dozen of gaseous debris were generated 
to move around it. The ring-like 
sky-distribution of the dwarf galaxies is just a
result of a view from inside of the Galaxy. We now
extend Kunkel's idea \citep{kun79} to an earlier and 
larger-scale dynamics related to the 
whole LGG structure. The data are due exclusively to
figures \ref{fig-1} to \ref{fig-4}.
 

An up-scaled model for the LGG dynamics
is as follows. Seen from the Galaxy, early 
still-extended gas-rich M31 encountered the early 
similar Galaxy at the perigalacticon of 
$\sim 150$ kpc about 10 Gyr ago. 
(These numerical values are to be determined through 
constructing the model quantitatively in the next section.)
The M31 orbit is an unclosed elliptical, and M31 leaves the 
Galaxy until it reaches the apogalacticon at
$\sim 1$ Mpc about 4 Gyr ago and then starts to 
approach the Galaxy.

At this early off-center collision, the still-extended 
gas of the Galaxy and that of M31 compresses each other 
hydrodynamically to generate a high density region of gas 
at a half way 
between them. A number of gas clouds are driven to
condense and scattered on the orbital 
plane. Some tens of them evolved into dwarf galaxies 
as we now observe them.
Most of these dwarfs gather gravitationally around the Galaxy or M31,
and other several ones are left behind from 
such groupings, belonging neither to the Galaxy nor to 
M31. In our model, the
orbital plane of the Galaxy-M31 is assumed to be identical
with the LGG disk, determined in figure \ref{fig-3}, or with
the plane defined by the rings in figures \ref{fig-1} 
and \ref{fig-2}.

In this connection, we refer to \citet{dee98},
in which the violently merging gas-rich galaxies
are observed just like shedding out a number of 
young dwarf galaxies around them. These authors claim that 
a considerable portion of the observed dwarf galaxies
are not cosmological objects but recently born ones in
and around the interacting galaxies. A recent HST 
observation by \citet{wea01} reveals clearly
that the disk galaxy NGC7318 in the 
Stepfan's Quintet is generating dozens of dwarf galaxies
in tidal interaction with its nearby galaxy. These 
dwarfs appear to distribute dominantly on an orbital 
plane of NGC7318 and its counterpart. Our LGG model 
can be, therefore, regarded as an earlier and 
larger-scaled version of these interacting galaxies and 
newborn dwarfs around them.

\begin{figure*}
 \includegraphics[width=17cm]{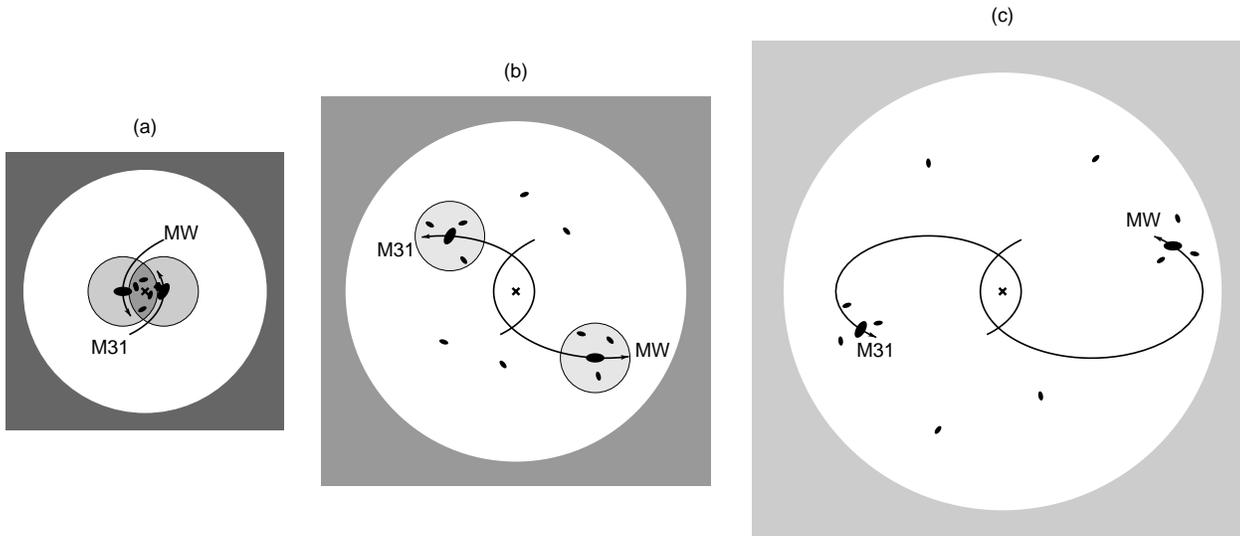}
\caption{Schematic snapshots of the dynamical 
evolution of the LGG members 
in local cosmic expansion. Figures \ref{fig-5}(a) to \ref{fig-5}(c) are 
face-on views of the LGG disk and one of which, figure \ref{fig-5}(c),
corresponds to figure \ref{fig-4}. 
A visual explanation is given for determining 
the orbits of M31 and the dwarf members. The radii of the LGG
spherical regions and the darkness around them represent
the expanding universe seen in local, and the cross 
denotes the center-of-mass of the Galaxy and M31. 
(a): 12 to 10 Gyr ago. 
The gas-rich proto-Galaxy and the similar proto-M31 
approach each other along their elongated orbits,
making an off-center collision to compress 
their halo gasses hydrodynamically. 
The early dwarfs, including the LMC and the SMC, are driven 
to form there. This high-density region is approximated 
geometrically to an overlapped area of the large halo 
spheres of the early Galaxy and M31. Since it is
colored in dark gray in figure \ref{fig-5}(a), it is referred to in 
the text as a {\it gray area}, and some times as
(halo) {\it overlapped region}, (gas) {\it compressed region}, 
(dwarf) {\it formation site}, etc.
 (b): $\sim 8$ Gyr ago. 
The newborn dwarf 
galaxies are scattered on the orbital plane of the Galaxy
and M31. After 4 Gyr of this epoch, the Galaxy and M31 pass their 
apogalacticons at a distance of 1 Mpc from each other, 
and start approaching. Some dwarf members 
are already partitioned into the neighborhoods of
the Galaxy and M31. 
(c): At present. M31 is at 770 kpc
from the Galaxy and approaching us with radial 
velocities of 120 km s$^{-1}$. This figure
suggests the existence of the azimuthal motion of
M31 relative to the Galaxy. 
}
\label{fig-5}
\end{figure*}

Figures \ref{fig-5}(a) to \ref{fig-5}(c)
are given to explain visually 
our model for the LGG origin and evolution. A spherical outer
surface of the LGG region and the darkness outside it mimic 
the expansion of the universe seen in the local coordinates. 
The mass in this sphere is assumed to have condensed exclusively to two 
mass-dominant galaxies, the Galaxy and M31, and two massive 
dwarfs, the LMC and the SMC.
Other dwarfs including M33 are treated as test
particles which move in time-dependent gravitational potential
due to the above four galaxies in the expanding sphere.

Since the radiation energy is far less than the rest mass
energy in our cosmological space and time, this spherical
surface of $R$ in radius follows \citep{lan75},
\begin{equation}
R=R_{0}(1-\cos \eta),
\label{eq4}
\end{equation}
and
\begin{equation}
t=t_{0}(\eta-\sin \eta),
\label{eq5}
\end{equation}
for closed universe, and
\begin{equation}
R=R_{0}(1-\cosh \eta),
\label{eq6}
\end{equation}
and
\begin{equation}
t=t_{0}(\eta-\sinh \eta),
\label{eq7}
\end{equation}
for open universe, and
\begin{equation}
R \propto t^{2/3}
\label{eq8}
\end{equation}
for flat universe, where $R_{0}$ and $t_{0}$ are
constants and $\eta$ is a parameter. 
In the present paper, we adopt
equation (\ref{eq8}), although choice of the universe is
not crucial on our discussion for the dynamics of the LGG
members. The dark energy that could accelerate the
Hubble expansion unnegligibly in this 4 Gyr is not
taken into account for simplicity. Also we ignore the distortion of
the spherical surface of the LGG region, because the
perturbing force due to the
gravitational quadruple moment of the LGG members decays
quickly as $1/R^{3}$.

In figure \ref{fig-5}(a) early gas-rich M31 and the
similar Galaxy approach each other, and subsequently 
hydorodynamic off-center collision occurred between them
to generate a high-density region of gas where a number of
gas clouds, which include the proto-LMC and the prot-SMC, 
form. The collision epoch is $\sim 10$ Gyr ago. This region is modeled in the present paper by
a geometrically overlapped area of the early-large-sized Galaxy and 
the similar M31. It will be referred to hereafter  
as an {\it overlapped region}, 
(hydrodynamically gas) {\it compressed region}, 
(dwarf) {\it formation site}, 
etc. In figures \ref{fig-5}(b) and 
\ref{fig-5}(c), the primordial dwarf 
galaxies scatter mostly on the orbital plane
of M31 and the Galaxy, and each evolves to the dwarf members 
as observed presently. 
More exactly they move in the gravitational potential 
due to the Galaxy, M31, the LMC and the SMC. The orbital 
plane of the Galaxy and M31 in figures \ref{fig-5}(a)
 to (c) is 
taken to be the same as this page, and the newborn 
dwarfs distribute approximately on this plane with small but 
finite thickness. In concluding this section, we note
a CDM universe which is assumed implicitly in our
model. According to this widely-accepted theory,
cosmologically large-scale structures are
considered to have grown from primordial
density-fluctuations of baryonic and dark
matters, whose square of amplitude is 
distributed in inverse law of the wave 
number. We can suppose naturally that
the corresponding random 
motion was superposed on the Hubble 
expansion flow,
and its amplitude is also in a power-law
function of the wave number. The LGG dynamics that we have modeled 
in the present
paper is consistent with such a 
CDM universe. That is, the
proto-Galaxy and proto-Andromeda
were formed from such baryonic
and dark matter media in various 
vortices. Their early off-center collision 
is considered to occur in a finite 
probability.

\section{Orbital Motions of M31 and the Galaxy Over the Past
12 Gyr, Determined from the LMC/SMC Dynamics}

We now construct a LGG model quantitatively 
following figures \ref{fig-5}(a) to (c). First we survey numerically 
past orbits of M31 by fixing 
its radial velocity as that in Table \ref{tab-1} and varying its 
tangential velocities which are still unknown. 
Then we choose most reasonable 
orbits which satisfy our criteria that the orbital plane lies 
within the LGG disk of finite thickness (figure \ref{fig-3}),
and M31 was close to the Galaxy 
at a cosmologically early epoch.
The latter criterion is required for the early collision between
primordial M31 and the primordial Galaxy, and it means that these
two galaxies had deviation from Hubble expansion law, as stated
above. Such picture of the universe is quite different from
those adopted in previous studies
\citep{mis85,pee93,lyn99} where all LGG members
are assumed to follow initially the Hubble expansion, starting
from a single region of finite but small size.

The off-center collision compresses the halo gas hydrodynamically to generate
a high density region at a half way between 
the primordial M31 and Galaxy at their closest
approach. Figure \ref{fig-5}(a) shows that primordial dwarfs,
including the primordial LMC and SMC, 
form in this region and scattered roughly
on the Galaxy-M31 orbital plane, in the same direction as 
the M31 motion seen from the Galaxy.

We adopt a reasonable assumption that 
the LMC/SMC thus formed have been in a binary 
state for the past $\sim $10 Gyr and produces the Magellanic Stream. 
In other words, the LMC/SMC structure is a
key phenomenon for determining the  
dynamical parameters of the Galaxy-M31 orbit and 
of the LGG model numerically.
Simultaneously if our model proves realistic, one of our basic questions
about the large orbital angular
momentum of the LMC/SMC is answered automatically.

The overall dynamical structure and evolution
for other dwarfs of the LGG will be discussed 
in the next section. See \citet{hod92} for the
details of M31, and \citet{wes97} for the 
dynamics and physics of the LMC/SMC system.


In applying the four-body dynamics to the Galaxy, 
M31, the LMC and the SMC, we adopt the following parameters. 
We assume that M31 and the Galaxy have a dark halo 
each with radius of 300 kpc and that the 
rotation curve is flat in
both galaxies. Then, the total gravitating mass of M31, 
including dark halo mass, is 
$M_\mathrm{A}=4 \times 10^{12}\MO$
for the rotation velocity $V=250$ km s$^{-1}$, and that of 
the Galaxy is $M_\mathrm{G}=3\times 10^{12}\MO$
for $V=220$ km s$^{-1}$.
Note that we are rediscussing the masses of M31 and the Galaxy
in section 7.

M31 is located at
770 kpc of the sun in the direction of $(l, b)$
=(121$^{\circ}, -22^{\circ}$). The heliocentric 
radial velocity of M31 is $-298$ km s$^{-1}$ 
\citep{tul88}, which corresponds to the 
radial velocity of $-120$ km s$^{-1}$ relative to 
the Galactic center. The masses of the LMC and the SMC are
taken to be 2$\times 10^{10}\MO$ and
2$\times 10^{9}\MO$, respectively, which have been 
applied in models by \citet{mur80} and 
\citet{gar94}.  
Numerical values for all other parameters are
conventional and the same with those applied in the
four-body dynamics in \citet{fuj99}
and \citet{saw99}.

We follow the orbits of
M31, the LMC, and the SMC (relative to the Galaxy) backward in time 
from present to the 
beginning of the universe, 10-13 Gyr ago, as schematically
shown in the ^^ ^^ expanding'' (or exactly, "shrinking," because we
follow the dynamics backward in time) LGG sphere in 
figures \ref{fig-5}(a) to \ref{fig-5}(c).
For calculation of the orbits, we assume various tangential
components of the velocity of M31 $(V_{l},V_{b})$ in 
the Galactic longitude and latitude. 
We take the following four conditions (i) to (iv) 
all of which should be satisfied in reasonable cases  
and also in order to restrict the values of $(V_{l}, V_{b})$ 
within some limited ranges beforehand. 
(i) The primordial M31 passes by 
the primordial Galaxy at their cosmologically early 
age, say, 9 to 13 Gyr ago.
(ii) The orbit of M31 is roughly on the plane determined in 
figure \ref{fig-3}. 
In this case the nodal 
line links the Galactic center and the present 
position of M31. (iii) M31 orbits the Galaxy
counterclockwise seen from the present position
of the sun, in the same sense as the LMC/SMC. 
(iv) The LMC and the SMC pass computationally through the
gray petal-like region located midway between the
primordial M31 and Galaxy in figure \ref{fig-5}(a).
The passage must be at the same time when this
hydrodynamically-compressed region is formed.

We apply the following equation (\ref{eq-9}) to our tracing 
back-in-time the Andromeda galaxy (M31), supplemented 
with other three similar equations (\ref{eq-10}) to (\ref{eq-12}) for the
Galaxy, the LMC and the SMC;

\begin{equation}
M_\mathrm{A}\hbox{\boldmath $\ddot r$}_\mathrm{A}=
-\frac{\partial \phi_\mathrm{G}}{\partial \hbox{{\boldmath $r$}}_\mathrm{A}}
-F_\mathrm{A}^\mathrm{G}\frac
{\hbox{\boldmath $\dot r$}_\mathrm{A}-\hbox{\boldmath $\dot r$}_\mathrm{G}}
{\left|\hbox{{\boldmath $\dot r$}}_\mathrm{A}-\hbox{{\boldmath $\dot r$}}_\mathrm{G}\right|},
\label{eq-9}
\end{equation}
\begin{equation}
M_\mathrm{G}\hbox{\boldmath $\ddot r$}_\mathrm{G}=
-\frac{\partial \phi_\mathrm{A}}{\partial \hbox{{\boldmath $r$}}_\mathrm{G}}
-F_\mathrm{G}^\mathrm{A}\frac
{\hbox{\boldmath $\dot r$}_\mathrm{G}-\hbox{\boldmath $\dot r$}_\mathrm{A}}
{\left|\hbox{{\boldmath $\dot r$}}_\mathrm{G}-\hbox{{\boldmath $\dot r$}}_\mathrm{A}\right|},
\label{eq-10}
\end{equation}
\begin{eqnarray}
M_\mathrm{L}\hbox{\boldmath $\ddot r$}_\mathrm{L} & = &
-\frac{\partial}{\partial \hbox{{\boldmath $r$}}_\mathrm{L}}
\left(\phi_\mathrm{A}+\phi_\mathrm{G}+\phi_\mathrm{S}\right) \nonumber \\
& & -  F_\mathrm{L}^\mathrm{A}\frac
{\hbox{\boldmath $\dot r$}_\mathrm{L}-\hbox{\boldmath $\dot r$}_\mathrm{A}}
{\left|\hbox{{\boldmath $\dot r$}}_\mathrm{L}-\hbox{{\boldmath $\dot r$}}_\mathrm{A}\right|}
-F_\mathrm{L}^\mathrm{G}
\frac
{\hbox{\boldmath $\dot r$}_\mathrm{L}-\hbox{\boldmath $\dot r$}_\mathrm{G}}
{\left|\hbox{{\boldmath $\dot r$}}_\mathrm{L}-\hbox{{\boldmath $\dot r$}}_\mathrm{G}\right|},
\label{eq-11}
\end{eqnarray}
and
\begin{eqnarray}
M_\mathrm{S}\hbox{\boldmath $\ddot r$}_\mathrm{S} & = &
-\frac{\partial}{\partial \hbox{{\boldmath $r$}}_\mathrm{S}}
\left(\phi_\mathrm{A}+\phi_\mathrm{G}+\phi_\mathrm{L}\right) \nonumber \\
& & -F_\mathrm{S}^\mathrm{A}
\frac
{\hbox{\boldmath $\dot r$}_\mathrm{S}-\hbox{\boldmath $\dot r$}_\mathrm{A}}
{\left|\hbox{{\boldmath $\dot r$}}_\mathrm{S}-\hbox{{\boldmath $\dot r$}}_\mathrm{A}\right|}
-F_\mathrm{S}^\mathrm{G}
\frac
{\hbox{\boldmath $\dot r$}_\mathrm{S}-\hbox{\boldmath $\dot r$}_\mathrm{G}}
{\left|\hbox{{\boldmath $\dot r$}}_\mathrm{S}-\hbox{{\boldmath $\dot r$}}_\mathrm{G}\right|},
\label{eq-12}
\end{eqnarray}
where the geometrical and dynamical symbols with suffices A, G, L and S
are referred to those of the Andromeda (M31), the Galaxy, the LMC
and the SMC, respectively. The gravitational potentials $\phi_{i}$ $(i={\rm A}$, 
G) are given by the potential for the flat rotation, 
and those $(i={\rm L}$ and S) are softened conventionally by introducing their central
cores of finite radii. 
The dynamical frictions, given by the
extremely-right ends of equations (\ref{eq-9}) to (\ref{eq-12})
are due primarily to the extended halos of the M31 and the
Galaxy. The two factors in equations (\ref{eq-11})
and (\ref{eq-12}), ($F_\mathrm{L}^\mathrm{A}$, $F_\mathrm{L}^\mathrm{G}$) 
and  ($F_\mathrm{S}^\mathrm{A}$, $F_\mathrm{S}^\mathrm{G}$), denote
the frictions on the LMC and the SMC due to the halos of M31 and the Galaxy. 
No frictions are considered between the LMC and the SMC. The adopted 
frictions $F_{i}^{j}$ $(i={\rm A}$, G, L, S, and $j={\rm A}$,
G) are as follows \citep{bin87}:
\begin{equation}
F_{i}^{j}
=-\frac
{ GM_{i}^{2} \ln \Lambda}
{
\left|\hbox{{\boldmath $r$}}_{i}-\hbox{{\boldmath $r$}}_{j}\right|^{2}
X_{j}^{2}
}
\left\{{\rm erf}(X_{j})-\frac{2X_{j}}{\sqrt{\pi}}\exp(-X_{j}^{2})
\right\}
\label{eq-13}
\end{equation}
where $X_{j}=\left|\hbox{{\boldmath $\dot r$}}_{i}-\hbox{{\boldmath $\dot r$}}_{j}\right|/V_{j0}$,
and $V_{j0}$ of $j={\rm A}$ and G are the flat rotation velocities of the Andromeda galaxy 
and the Galaxy.
The $\ln \Lambda$ is the Coulomb logarithm for the gravitatial force and we use
the value $\ln \Lambda =3.0$ in our simulations. The other notations denote
their usual meanings. 

\begin{figure*}
\resizebox{\hsize}{!}{\includegraphics{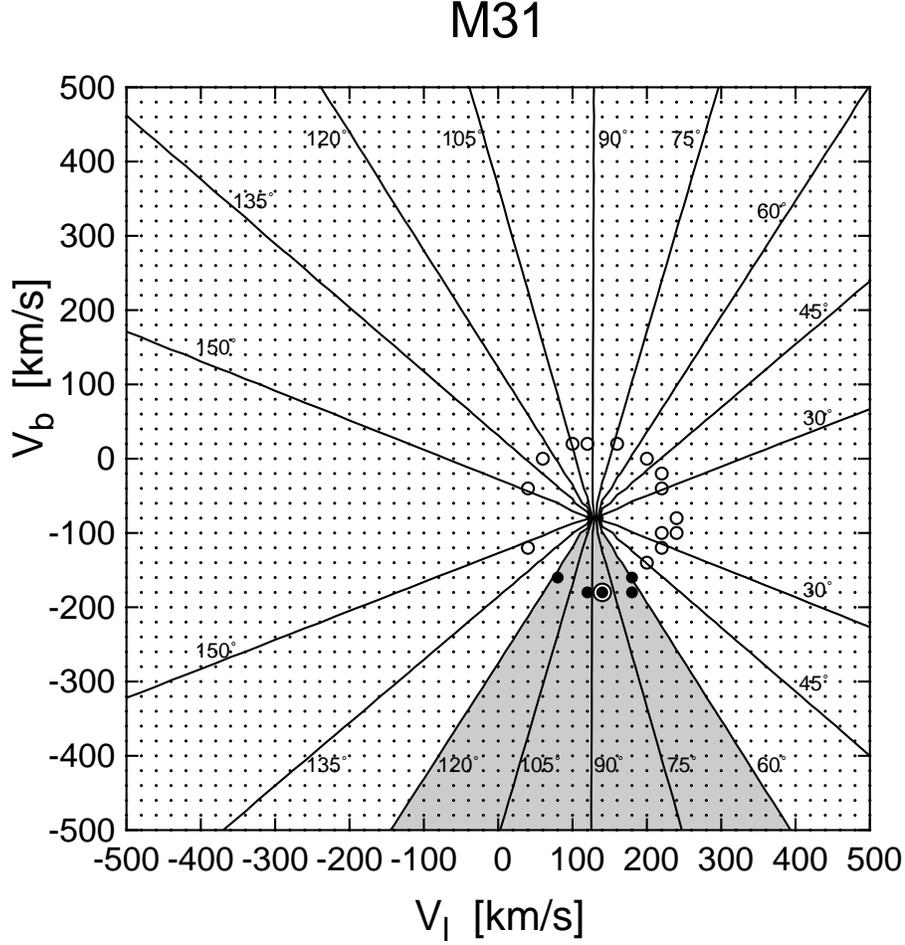}}
\caption{Tangential velocities of M31 $(V_{l},V_{b})$ assumed 
for tracing its past orbits.
Model orbits of M31, computed with the tangential 
velocities of open and filled circles, can 
reproduce the dynamics that the LMC/SMC system
was driven to form in the gray area in figure \ref{fig-5}(a),
and has been in a binary state for this $10^{10}$ yr, reproducing
the Magellanic Stream through particle simulations. 
Other velocities of small dots 
cannot reproduce these dynamics. 
Roughly-radial contour lines indicate the angle 
between the orbital angular momentum of M31 and the 
direction of the Galactic north pole ($z$ axis). 
If we apply the 
tangential velocities $(V_{l},V_{b})$ shown by
filled circles in the lower shaded region, the orbit
of M31 is in the LGG disk of finite thickness, and one of the most 
representative tangential velocities are at a double circle
$(V_{l},V_{b})=(140$ km s$^{-1}$, $-180$ km s$^{-1}$).
The corresponding proper motion is 
$(\mu _{l}, \mu _{b})$
=$(38~\mu {\rm as~yr}^{-1}, -49~\mu {\rm as~yr}^{-1})$.
}
\label{fig-6}
\end{figure*}

Figure \ref{fig-6} shows the 
tangential velocities $(V_{l},V_{b})$, 
in the extensive range of 
$-500$ km s$^{-1} \le V_{l} \le 500$ km s$^{-1}$ 
and $-500$ km s$^{-1} \le V_{b} \le 500$ km s$^{-1}$,
adopted in our backward search of the M31 orbits 
that satisfy the above four conditions (i) to (iv).
The orbits, computed from the 
tangential velocities in the upper- and lower-opened 
wedges of $(V_{l},V_{b})$, are confined within 
$\pm 30^{\circ}$ about
the plane normal to the line joining the Galactic center
and the present position of the sun, and then the nodal line
links the Galactic center and the present
position of M31. In particular, the M31 orbits in the 
lower wedge of figure \ref{fig-6} satisfy the conditions (ii) and (iii),
and those of the filled circles produce the model 
orbits which pass the peri-Galacticon of 140 kpc about 
10 Gyr ago at the same time when the Magellanic 
Clouds System goes through just this midway between M31
and the Galaxy. This dynamics is consistent with 
the condition (iv). It is easily seen that if
the conditions (ii) to (iv) are satisfied,
the condition (i) is automatically so.

As mentioned in the caption of figure \ref{fig-6}, we can choose
the tangential velocity $(V_{l},V_{b})=(140$ km s$^{-1}$, $-180$ km s$^{-1}$)
marked with double circle as the most probable or the most realistic 
value among five filled circles.
The corresponding proper motion is, if observed in near future,
$(\mu _{l}, \mu _{b})$
=$(38\ \mu {\rm as~yr}^{-1}, -49\ \mu {\rm as~yr}^{-1})$.
It is noted that 
the conversing point of the radial lines 
$(V_{l},V_{b})=(130$ km s$^{-1}$, $-80$ km s$^{-1}$) 
in figure \ref{fig-6} is 
due to the vectorial summation of the Galactic
 rotation and the solar motion.

\begin{figure*}
\begin{center}
\includegraphics[width=15cm]{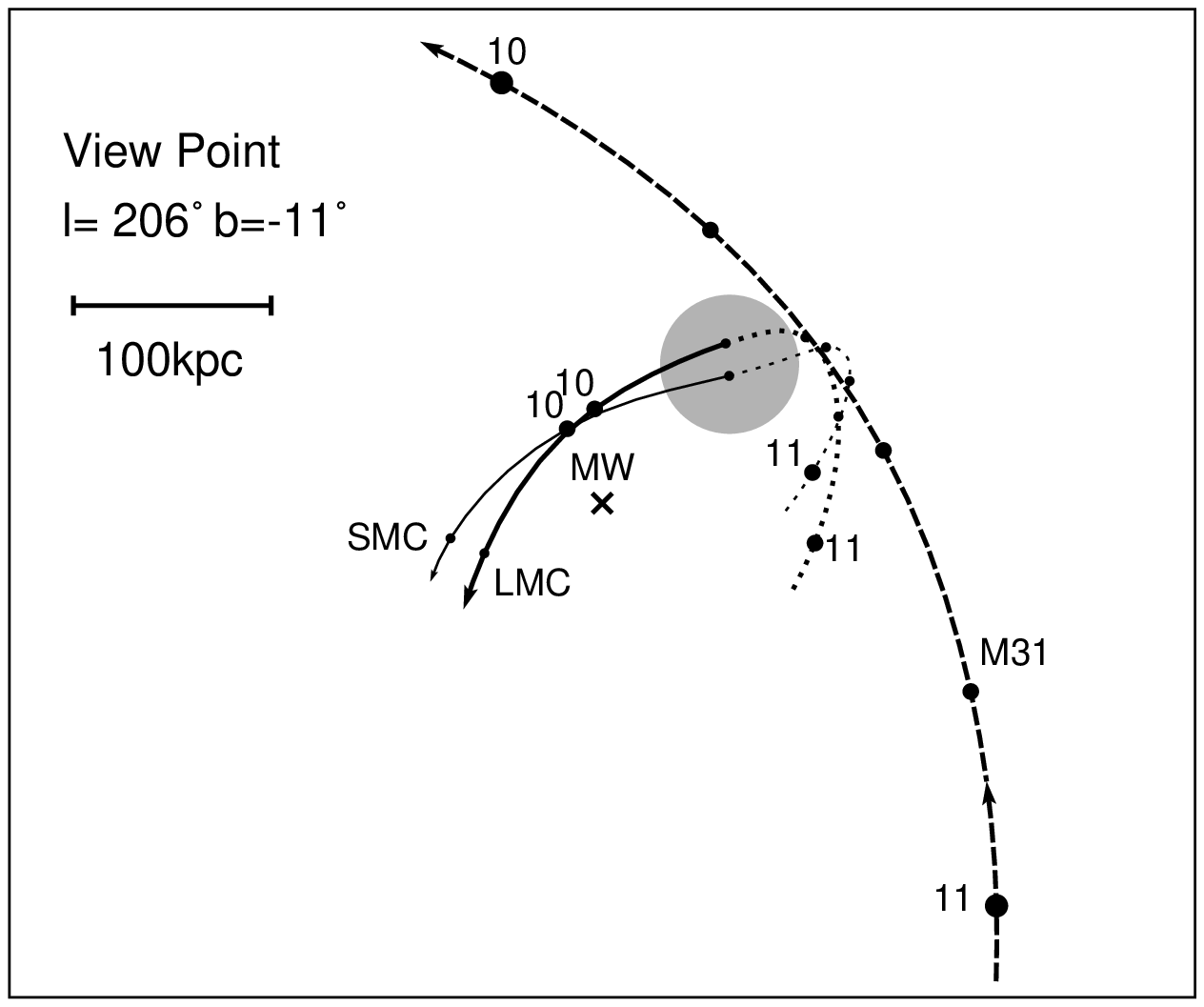}
\caption{Time-, position- and direction-matching
between the cosmologically early orbits of 
M31 and that of the LMC/SMC
on the LGG plane. It is viewed from the
same direction of figure \ref{fig-4} or $(l,b)=(206^{\circ},-11^{\circ})$. 
They are given relative to the Galaxy (the cross at center). 
In order to avoid 
confusion, the orbits later than 10 Gyr are not given.
The orbital planes of M31 and the 
LMC/SMC are nearly on this page. The arrows 
indicate the direction of motion toward
the present and the attached 
numerals indicate the time in 1 Gyr measured from the present. 
The gray region denotes a formation site of 
the LMC/SMC system. Since the orbits of the LMC/SMC before 
entering this region do not concern the present model, we 
show them only in dotted lines in order to see how they are
matched with the M31 orbit in position, time and velocity.}
\label{fig-7}
\end{center}
\end{figure*}


The most probable model orbits, projected onto 
the LGG plane are shown in figure \ref{fig-7}.
There can be seen that the orbital plane of M31 coincides
approximately with that of the LMC/SMC system, and M31
makes an off-center collision with the Galaxy. 
A large-scale compression of primordial gas seems to be realized.  
The epoch of the collision is determined as 10.4 Gyr ago. 
Moreover importantly, the LMC/SMC also pass across this region  
10.4 Gyr ago; the motions of M31 and the LMC/SMC 
thus lend a support to our model idea that
primordial dwarfs, including the LMC and the SMC, were
driven to form in the high density region of gas.
Only a small portion of the orbital
angular momentum of M31 (around the Galaxy) is 
transferred to those of the LMC/SMC simultaneously.

\begin{figure*}
\includegraphics[width=17cm]{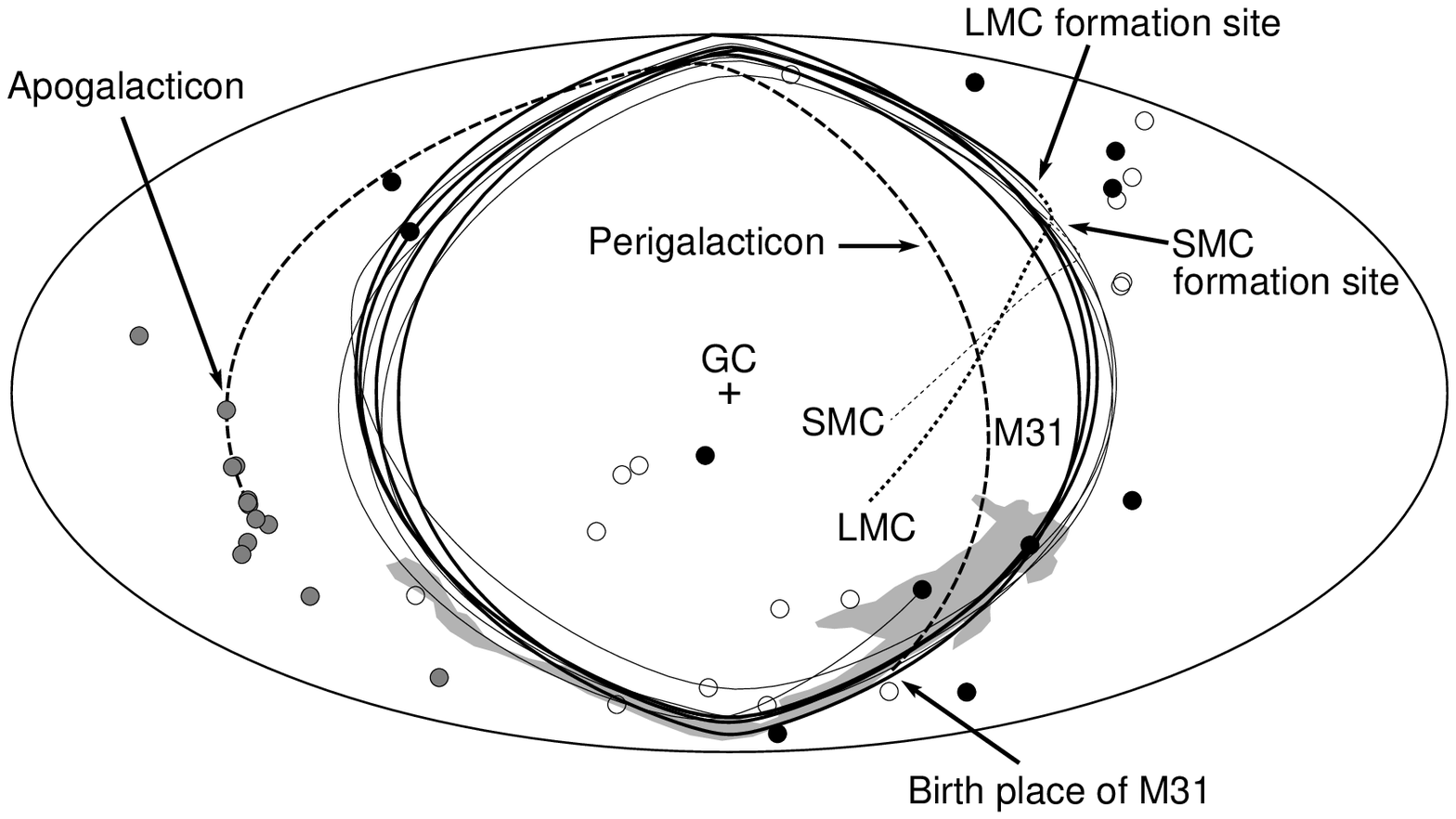}
\caption{Orbits of M31 and the LMC/SMC for the last 10.4 Gyr. 
Since they are 
projected onto the $(l,b)$-plane, this figure is 
topologically the same as figure \ref{fig-7}. 
The formation site of the LMC/SMC  
is in the direction of
$(l,b)=(278^{\circ},45^{\circ})$. 
The extreme end of the dashed line 
is a birth place of M31, estimated in the flat universe of 
$H_{0}=75 {\rm ~km\ s}^{-1} {\rm Mpc}^{-1}$. 
}
\label{fig-8}
\end{figure*}

The orbits of M31, the LMC and the SMC over the past 
12 Gyr are shown in figure \ref{fig-8}, 
projected onto the $(l, b)$ plane. 
Note that the orbits of the LMC and the SMC earlier
than 10.4 Gyr do not concern our model construction.
A comparison with 
figures \ref{fig-1} and \ref{fig-2}
presents actually a two-dimensional coincidence 
between the model orbits and the rings or circles
which trace the sky distribution of the LGG members.
The perigalacticon of the M31 orbit is in the direction of
$(l, b)= (307^{\circ},34^{\circ})$, and the
apogalacticon in (119$^{\circ},-1^{\circ}$). Their 
distances, measured from the Galactic center,
are respectively 140 kpc and 980 kpc. 
The formation site of the LMC/SMC is near in the 
direction of the 
perigalacticon. Of course, this region is also the 
formation site of the LGG dwarfs as shown in gray 
in figure \ref{fig-5}(a).

The orbits and attached numerals in 
figures \ref{fig-7} and \ref{fig-8} 
are more-or-less variable depending on
the numerical values for the 
parameters of the LGG and the age of the universe. 
However, so far as they change only within some reasonable 
ranges, the overall features of the orbits of M31 and 
those of the LMC and the SMC remain essentially unchanged. For future
studies of the LGG, therefore, it would be useful to
predict again here numerically the proper motions of M31,
\begin{equation} 
(\mu _{l},\ \mu_{b})
              =(38 \pm 16 \ \mu{\rm as~yr}^{-1}, 
               -49 \pm  5 \ \mu{\rm as~yr}^{-1}).
\end{equation}
\noindent
Here we may not understand why the proper motion 
$\mu _{l}$ is comparable to $\mu_{b}$ in magnitude,
because the orbital plane of M31 has been determined as 
nearly perpendicular to 
the Galactic plane. To this question, we recall that 
the transverse velocity to be observed is contributed 
from the Galactic rotation of 220 km s$^{-1}$,
directed off from M31 by more than 30$^{\circ}$.

\section{Origin and Dynamics of Dwarf Members of the LGG}

In our LGG model, the dwarf members, 
including the LMC and the SMC, are driven to form in the
high-density region of gas generated by the ancient
off-center collision between M31 and the Galaxy. 
Newborn dwarf-galaxies are scattered approximately
on the orbital 
plane of these two massive galaxies, and move
in the time-dependent potential due mostly to 
the Galaxy and M31. The grown-up dwarfs tend 
to gather around either the 
Galaxy or M31, with minor ones left behind 
from such groupings. They are observed in the space 
between M31 and the Galaxy as if connecting geometrically
these two galaxies (figures \ref{fig-3} and \ref{fig-4}).

In order to check the validity of our model, we
here examine whether or not the dwarf 
members in figure \ref{fig-3} and Table \ref{tab-1} have their past orbits 
which are consistent with our model scenario. 

Exactly in the same way as for the case of the
past orbits of the LMC and the SMC, we trace back in time 
the dwarfs' orbits from their present positions, 
with observed radial velocities (Table \ref{tab-1})
and tangential velocities assumed on the 
$(V_{l},V_{b})$ plane as in figure \ref{fig-6}.
Again the time-dependent 
gravitational potential due to the Galaxy, M31,
the LMC and the SMC is adopted, and 
the latter two objects are required to be in 
a binary state for the past $10^{10}$ yr and to reproduce
the geometrical and dynamical structures of the 
Magellanic Stream through the particle simulation.

We have thus surveyed computationally 
all past orbits of the dwarf members  
in Table \ref{tab-1}, and examined
if they pass across the overlapped region of the
primordial Galaxy and the primordial M31
$\sim $10.4 Gyr ago (figure \ref{fig-7}).
Note that the dwarfs with no available 
data on radial velocities in Table \ref{tab-1}
are excluded from our orbit computations.

\begin{figure*}
\begin{center}
\includegraphics[width=12cm]{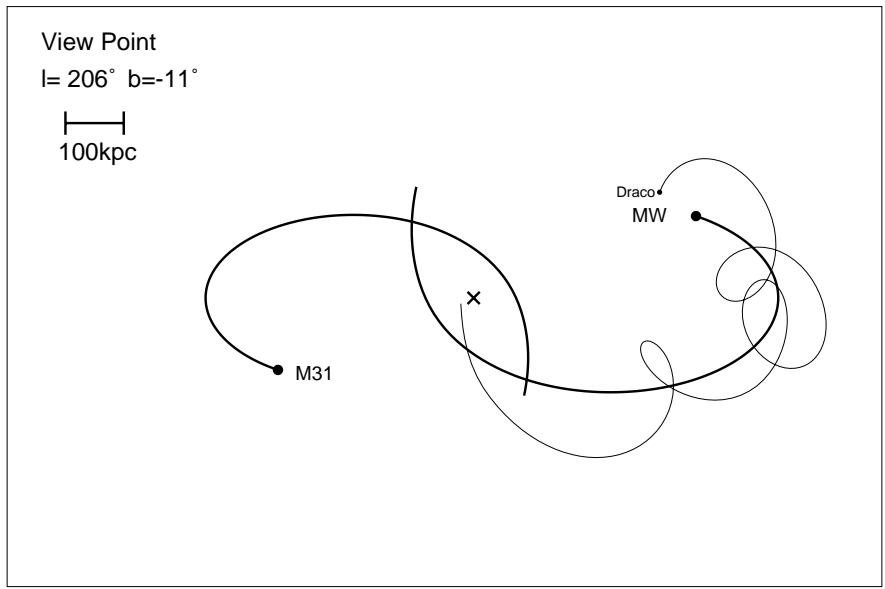}
\caption{Orbit of the dwarf spheroidal galaxy (Dsph)
Draco projected onto the LGG plane. The formation 
site is the same as in figure \ref{fig-7}. 
}
\label{fig-9}
\end{center}
\end{figure*}

Figure \ref{fig-9} shows an orbit of the LGG dwarf spheroidal,
Draco (Dsph), which is presently at $(l, b)=(86.4^{\circ},34.7^{\circ})$ 
and 82 kpc distant from the sun (Table \ref{tab-1}). We can see 
that the orbit starts from the same formation site as 
that for the LMC/SMC, and arrives at the present position 
of the Draco. 

\begin{figure*}
\resizebox{\hsize}{!}{\includegraphics{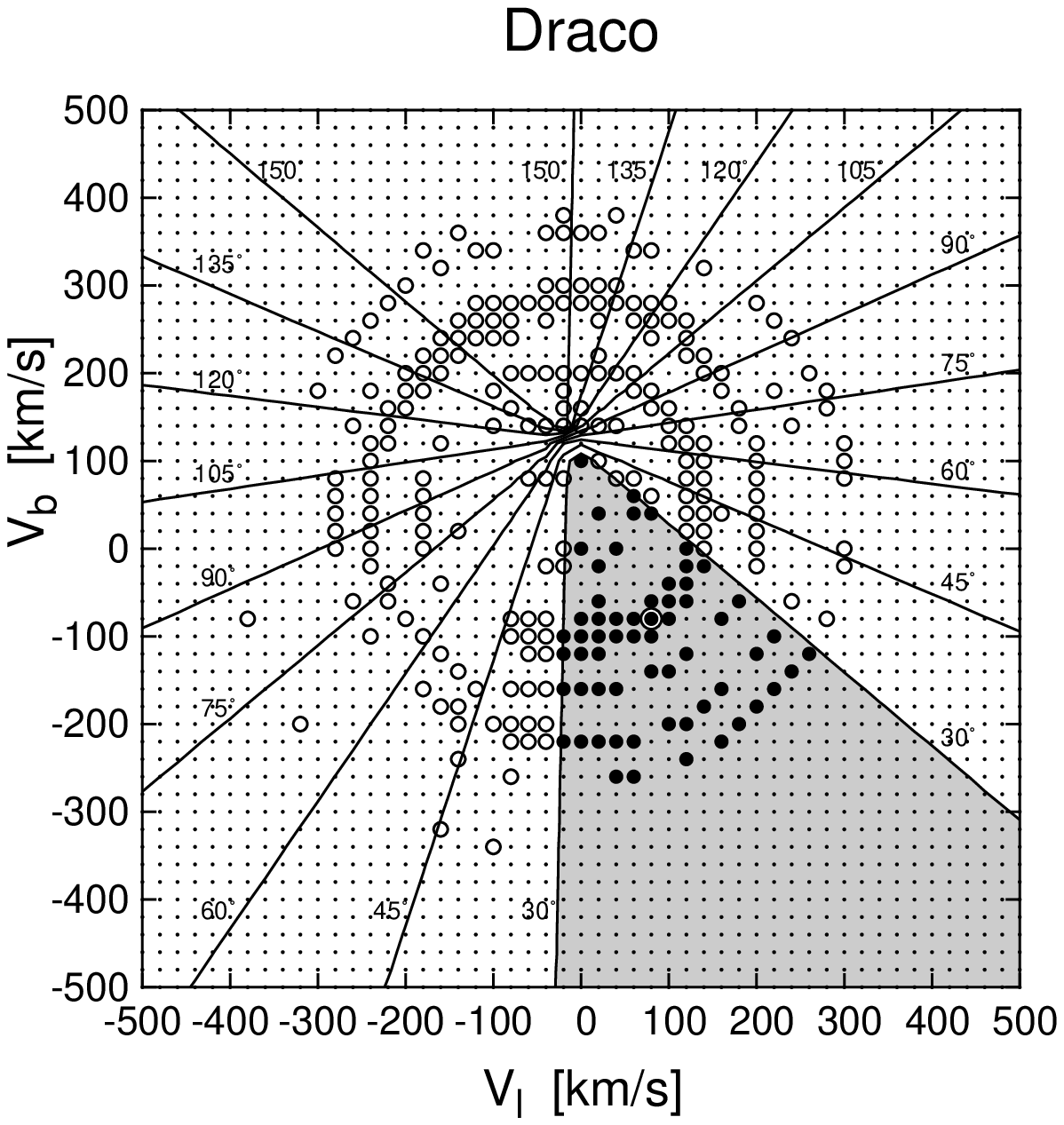}}
\caption{Velocity diagram $(V_{l},V_{b})$ of
the tangential motion, prepared for tracing 
back-in-time the Draco's 
orbit until its birth place. If we choose 
velocities in the shaded region, 
the Draco's motion is confined within the LGG 
disk of finite thickness
of 100 kpc, and moves in the same sense
as driven hydrodynamically by the collision with 
M31. The orbit in figure \ref{fig-9} is obtained from the
tangential velocity at the double circle. }
\label{fig-10}
\end{figure*}

\begin{table*}[t]
\caption{A sheet of score for model orbits and their 
proper motions to be measured, if possible, at present.
The LGG members whose orbits agree with our model scenario
have a mark $\bigcirc$, and those that do not agree
$\times $. The proper motion $\mu_{l}$ and $\mu_{b}$ 
of each galaxy with $\bigcirc $ is given in 
$\mu$as yr$^{-1}$ in the last two columns.
\label{tab-2}
}

\begin{center}
\begin{tabular}{lcrrclcrr} \hline \hline
Galaxy             &score      &$\mu_{l}$&$\mu_{b}$&&   Galaxy       & score &$\mu_{l}$&$\mu_{b}$ \\ \hline
WLM 		   &$\times$   &---&---&&       Leo A 	   &$\times$   &---&---\\
NGC55 		   &$\times$   &---&---&&       Sextans B  &$\times$   &---&---\\
IC10  		   &$\bigcirc$ &41 &$-61$&&     NGC3109    &$\times$   &---&---\\
NGC147 		   &$\bigcirc$ &35 & 12  &&     Antlia 	   &$\times$   &---&---\\
And III 	   &---        &---&---&&       Leo I 	   &$\bigcirc$ &$-17$&$-152$\\
NGC185 		   &$\bigcirc$ &41 &20 &&       Sextans A  &$\times$   &---&---\\
NGC205 		   &$\bigcirc$ &31 &$-88$&&     Sextans    &$\bigcirc$ &$-49$&$245$\\
M32 		   &$\bigcirc$ &37&$-115$&&     Leo II 	   &$\bigcirc$ &$21$&$-93$\\
M31 		   &$\bigcirc$ &38 &$-49$&&     GR8 	   &$\times$   &---&---\\
And I 		   &---        &---&---&&       Ursa Minor &$\bigcirc$ &256&$-128$\\
SMC  		   &$\bigcirc$ &$-730$&1360&&   Draco 	   &$\bigcirc$ &$206$&$-206$\\
Sculptor	   &$\bigcirc$ &$-347$&1090&&   Milky Way  &$\bigcirc$ &---&---\\
LGS3 		   &$\bigcirc$ &89 &$-130$&&    Sagittarius&$\bigcirc$ &$-3520$&1850\\
IC1613 		   &$\bigcirc$ &72 &$-36$&&     SagDIG 	   &$\times$   &---&---\\
And II 		   &---        &---&---&&       NGC6822    &$\times$   &---&---\\
M33 		   &$\bigcirc$ &90 &$-80$&&     DDO 210    &$\times$   &---&---\\
Phoenix 	   &$\bigcirc$ &$-9$&180&&      IC5152 	   &$\times$   &---&---\\
Fornax 		   &$\bigcirc$ &275&474&&       Tucana 	   &---        &---&---\\
EGB0427+63 	   &$\times$   &---&---&&       UKS2323-326&$\times$   &---&---\\
LMC 		   &$\bigcirc$ &$-420$&1720&&   Pegasus    &$\bigcirc$ &$-31$&$-57$\\
Carina 		   &$\bigcirc$ &$20$&522&&                 &           &&\\
\hline\hline
\end{tabular}
\end{center}

\end{table*}

The orbit has been determined
by assuming a proper motion on the
 $(V_{l},V_{b})$ plane in figure \ref{fig-10},
where each symbol has the same meaning 
as that in figure \ref{fig-6}, but nearly radial 
contour-lines indicate that $(V_{l}, V_{b})$ 
on these lines guarantee respectively same angles 
between the orbital plane of the Draco and that of
of the Galaxy-M31 system. The present proper motion
$(\mu _{l},\ \mu_{b})$ predicted for the Draco in Table \ref{tab-2},
is based on the orbit determined with the velocity $(V_{l},V_{b})$
at the double circle in figure \ref{fig-10}.


Similar orbit-surveys are carried out for the 
other dwarf members of the Galaxy group, 
and the results are given in Table \ref{tab-2}; open circles 
and crosses given in the second and sixth columns represent
respectively the existence and absence of the 
model-consistent orbits for the past 10.4 Gyr. Although
these results are discussed later on the basis of more
global model-picture, we can say here, at least, more
than sixty percent of the Galaxy group members hold $\bigcirc $.

The Sagittarius dwarf
galaxy, which was discovered by \citet{iba94}, is
one of the four dwarfs not on the LGG ring 
 (see figures \ref{fig-1} and \ref{fig-2}) but near the 
direction of the Galactic center. It 
locates at only 16 kpc of the Galactic center and is
elongated at a right angle against the Galactic plane.
Such elongated structure of the Sagittarius is considered to represent a 
tidal disruption due to the Galaxy and its
major axis should coincide approximately with the orbit
round the Galaxy. 
Thus we can estimate the orbital plane of the Sagittarius by
tracing its major axis and the Galactic center 
\citep{iba97}. 
Using the estimated orbital plane together with the 
data in Table \ref{tab-1} and our model conditions 
for the LGG origin, we have computed the model orbit 
of the Sagittarius dwarf galaxy, and obtained the result of 
$\bigcirc $ and the prediction for proper motion in the 
(l, b) directions in Table \ref{tab-2}. It is qualitatively 
consistent with that given in \citet{iba01}.

\begin{figure*}
\begin{center}
\includegraphics[width=12cm]{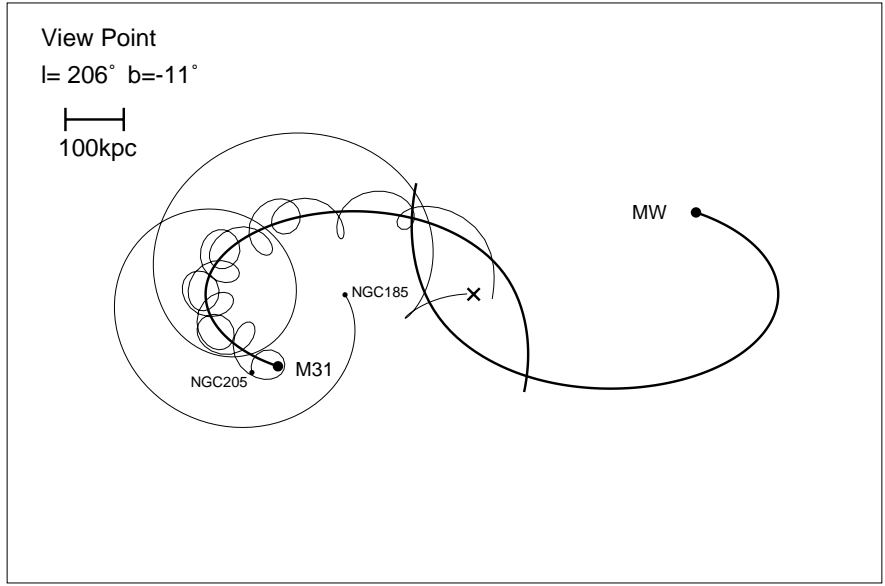}
\caption{Same as figure \ref{fig-9}, but the orbits of 
NGC185 and NGC205 of the M31 group. }
\label{fig-11}
\end{center}
\end{figure*}

Next we searched for model-fitting orbits of the M31 group
dwarfs in the same way as in cases of the Galaxy 
group. Two orbits of NGC185 and NGC205 thus obtained are presented 
in figure \ref{fig-11} over their lifetime of 10.4 Gyr. 
They are actually confined within the flat disk 
of finite thickness (figure \ref{fig-3}), and fulfill our model 
conditions that they pass 
the formation site exactly 10.4 Gyr ago. 
 
Fourteen dwarf galaxies out of thirty nine members (Table \ref{tab-1})
belong neither to the Galaxy nor to M31, but are
located just like linking these two massive galaxies (figure \ref{fig-3}). 
One of 
them, Pegasus, has orbit that follows our model scenario. 
A particular interest may also be in examining whether
or not our model orbits of NGC205 and M32 are 
consistent with those determined by \citet{sat86}
and \citet{byr76} who claimed that the warp of 
hydrogen gas distribution at the outer parts of M31
and the peculiar motion of hydrogen gas near the center
of M31 are due to the tidal interaction with NGC205
and M32, respectively. According to \citet{sat86}
the orbital plane of NGC205 makes a large angle of 
$\sim 90^{\circ}$ against the M31 disk and the 
direction of motion is counterclockwise seen from 
(l, b)$=(206^{\circ}, -11^{\circ})$.
These orbit features are consistent with 
in figure \ref{fig-11}.
Similarly, the orbital plane of M32 makes
a medium angle against the M31 disk and the sense of 
revolution around M31 \citep{byr76} are 
consistent with that inferred 
from the kinematics of our off-center collision
between M31 and the Galaxy.   

We mention that the model orbits for these two galaxies are 
determined 
easily, because NGC205 and M32 
are so close to M31, 20 to 40 kpc from its center, 
and, therefore, they are located already deep 
within the LGG disk of 50 to 100 kpc thickness. Since the 
scale of the formation site of $\sim 50-100$ kpc 
is much larger than the above mentioned 20 to 40 kpc, 
we can choose their suitable proper motion
in considerably wide ranges of the 
$(V_{l},\ V_{b})$ diagram as in figure \ref{fig-10}.

\citet{iba01} and \citet{irw01} determined the 
orbit of M32 by use of the 
data of stellar streaming in M31. It is shown easily 
to be consistent with our model orbit.
At present, however, we 
do not have enough data to conclude which is 
more agreeable to our model, Byrd's orbit \citep{byr76} or Ibata's 
one \citep{iba01}.    

Again we give in Table \ref{tab-2} the results, 
$\bigcirc $ and $\times $, for our orbit search for
the members of the M31 group, together with the results
for the nongroup members. 

\begin{figure*}
\includegraphics[width=17cm]{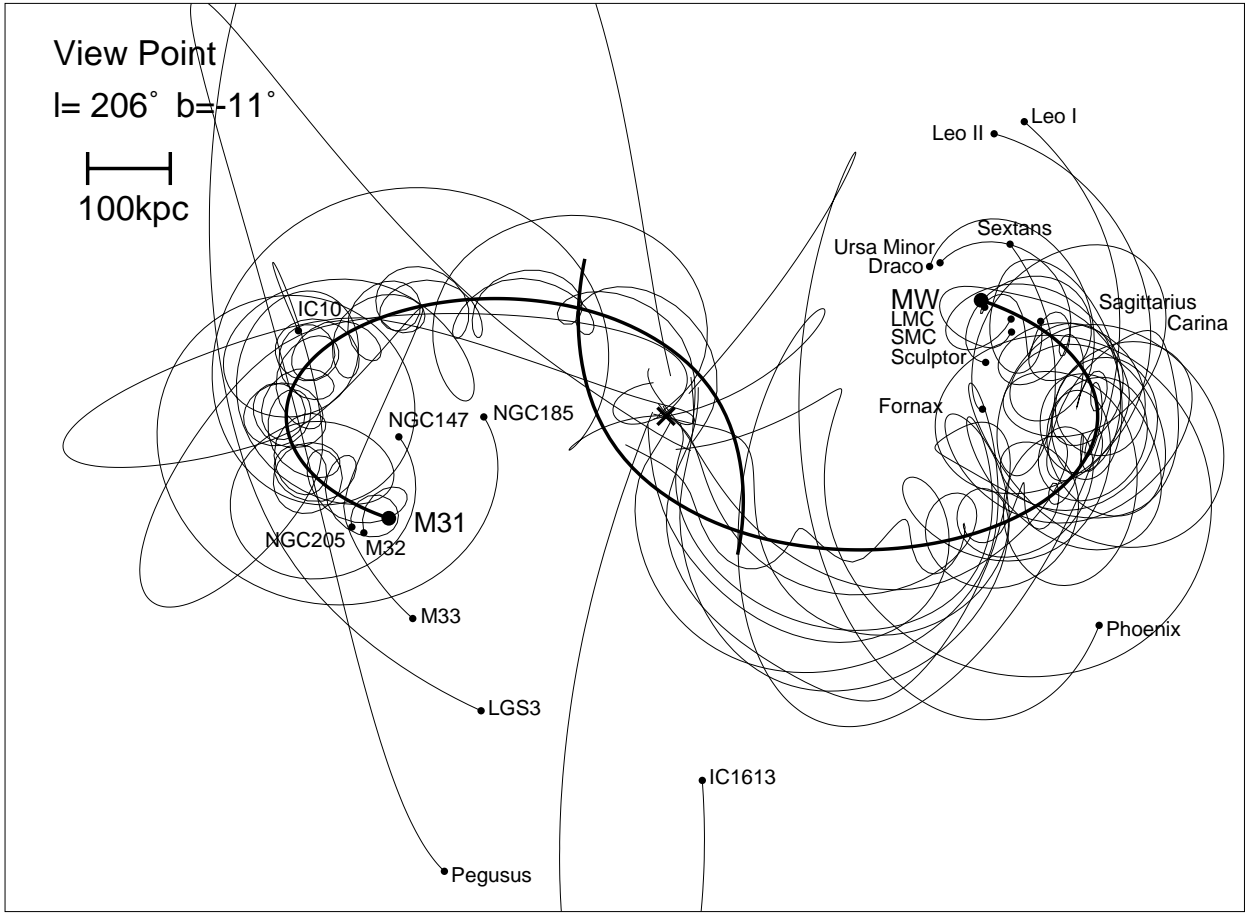}
\caption{Face-on view of orbits of the LGG 
members, projected of the orbital
plane of M31 and the Galaxy. All dwarf members 
in this diagram start from their common formation site 
and arrive at their present positions as we see them today.
The view is in the direction from 
$(l,\ b)=(206^{\circ},-11^{\circ})$.}
\label{fig-12}
\end{figure*}

\begin{figure*}
 \includegraphics[width=17cm]{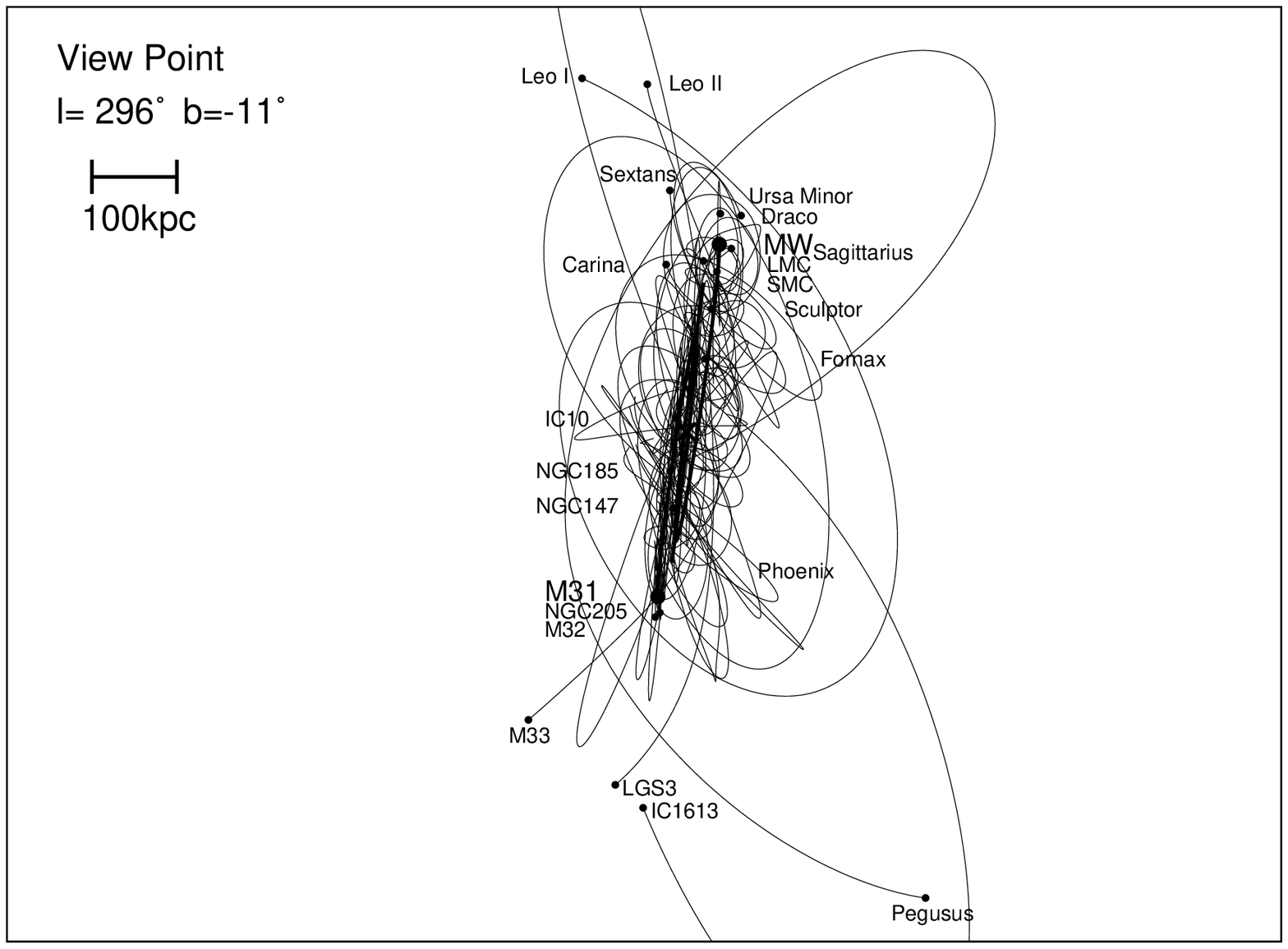}
\caption{Same as in figure \ref{fig-12}, but an 
edge-on view. Most of orbits are contained in the LGG
disk of finite thickness. The view is in the direction 
from $(l, b)=(296^{\circ},-11^{\circ})$. }
\label{fig-13}
\end{figure*}

Finally we show in figures \ref{fig-12} and \ref{fig-13} 
all orbits of the 
members of the LGG with $\bigcirc $ in Table \ref{tab-2}:
Figure \ref{fig-12} gives a face-on view seen from the direction of 
$(l, b)=(206^{\circ},-11^{\circ})$ while figure \ref{fig-13}
an approximately edge-on view from 
the direction of $(296^{\circ},-11^{\circ})$. 
Although the orbits
are much tangled each other and it is very hard
to derive their detailed numerical data, we can see globally
that the LGG dwarf galaxies are driven to form
by the ancient off-center collision between M31 and 
the Galaxy, and that they evolve dynamically
on the orbital plane of these two massive galaxies.

\section{Angular Momentum and Dark Matter Problems
in the LGG Model}\label{s-6}

We have made a hydrodynamical model for the
origin of the orbital angular momentum of
the Magellanic Cloud System, and also for the 
hydrodynamically-driven 
formation of the dwarf members of the LGG.
If our model is realistic,
the orbits of the LGG members and their 
formation sites can be determined theoretically.
However, we still have some problems that should 
be discussed from 
the viewpoint of the present model.

\subsection{Tidal torque on the primordial disks of 
the Galaxy and M31}
 
At the close passage of M31 at $\sim $150 kpc
from the Galactic center (figure \ref{fig-5}(a)), the 
tidal torque $N$ works on the 
disk, which could spin it up in the same
direction as that of the orbital angular 
momentum of the Magellanic Cloud System about
the Galaxy. It is perpendicular to 
the present disk of the Galaxy. This
dynamics should be exert the same effect 
on the disk of M31, too.
  
The gravitational torque is the monopole-quadruple 
moment interaction. That is,
$N\propto M/d^{3}$, where $M$ is the mass of 
the nearby passing galaxy and $d$ is the
impact parameter or the distance of the
perigalacticon. Note that $N$ is not
proportional to $d^{-2}$ but to $d^{-3}$.

We can estimate how much a spin has been 
given to the Galactic disk through the
off-center collision with M31.
Here we use conveniently the dynamics of the 
warping of hydrogen gas disk of the 
Galaxy that was modeled theoretically
by the close approach of the Magellanic 
Cloud System at $\sim $15 kpc of the 
Galactic center \citep{fuj76, fuj77}.
The torque, $N_{\rm LMC}$ due to the LMC, is,
\begin{equation}
N_{\rm LMC}\propto M_{\rm LMC}/
       \left(15 {\rm ~kpc}\right)^{3}.
       \label{eq-15}
\end{equation}
In the same way, the torque
due to M31, $N_{\rm M31}$, is represented as,
\begin{equation}
N_{\rm M31}\propto M_{\rm M31}/\left(150 {\rm ~kpc}\right)^{3}.
       \label{eq-16}
\end{equation}
The ratio of $N_{\rm M31}$ to $N_{\rm LMC}$ is, 
\begin{eqnarray}
{N_{\rm M31} \over {N_{\rm LMC}}} & = &
        \left({M_{\rm M31} \over M_{\rm LMC}}\right)
       \left({15 {\rm ~kpc} \over
       150 {\rm ~kpc}}\right)^{3} \nonumber \\
& \sim & {4\times 10^{12}M_{\odot} 
       \over 2\times 10^{10}M_{\odot}}
       \left({1 \over 10}\right)^{3}=0.2.
       \label{eq-17}
\end{eqnarray}

The angular momentum of 
the warp of hydrogen gas is known 
to be only a small fraction of that of 
the disk, and it is negligibly 
small for the inner disk of 8 kpc in radius. Thus the
ratio in equation (\ref{eq-17}) indicates that the
torque due to M31, $N_{\rm M31}$, is smaller than
$N_{\rm LMC}$, or our off-center collision
between M31 and the Galaxy is not large
to spin up the disk so that it becomes parallel
to the orbital plane of the Galaxy-M31
system. Even if we take into account the 
duration of these torques, the spin of 
the disk is still small, only comparable 
to that of the observed warping of the disk 
\citep{fuj76, fuj77}.

The ratio in equation (\ref{eq-17}) has been obtained 
for the disk whose radius was $\sim $15 kpc
at the time of the off-center collision.
If such a radius of the early disk 
was $\sim 75$ kpc, the magnitude of the quadruple 
moment becomes five times and, therefore,
the ratio in equation (\ref{eq-17}) is roughly
unity,
\begin{equation}
      {N_{\rm M31} \over {N_{\rm LMC}}}\sim 1. 
\end{equation}
The gain of the angular momenta was still 
small for the early large Galactic disk. Again it would
be only comparable to 
that of the present-day warp of the 
hydrogen disk. 

We can thus infer from these 
data that the orbital angular momentum of 
the Magellanic Cloud System and the spin of the 
Galactic disk are of the different origin.
These arguments may reflect the fact
that the Galactic disk and the M31 disk 
are observed not parallel on the plane of the 
sky but make an angle of about 60$^{\circ}$
against each other.

\citet{pee69} and \citet{thu77}
presented a tidal model that the gravitational 
torque due to the nearby passing M31 can spin up 
the disk of the Galaxy. 
However, the model parameter values and basic assumptions adopted there
are very different from those used in this study: The epoch of the 
interaction between the Galaxy and M31
is 12-13 Gyr ago; the M31's orbit is straight, 
passing by the Galaxy and changing its direction in the
neighborhood; the proto-Galaxy is already
in an elongated flat disk with 100 kpc length 
directed towards M31. 

\subsection{Dark matter in dwarf members}

Some of the LGG dwarfs have large amounts of
dark matter, inferred from their high 
mass-to-light
ratios of Draco, Sextans and Ulsa Minor
\citep{van00},
although it is still uncertain that
many other dwarfs do not have dark matter.
The present model for the LGG origin gives 
no answer to the question; why dark matter 
is not distributed equitably to all dwarf 
members of the LGG.

We present here a possible 
explanation by using figure 14 in which 
halo models of the proto-Galaxy and 
proto-Andromeda are given at the time 
just before encountering or grazing each 
other (figure \ref{fig-5}(a)), with a relative 
velocity of $\sim 400$ km s$^{-1}$.

The halo comprises baryonic matter clouds
and dark matter clouds, some of which
occupy the same space while others
do not. The mean random velocity of the
clouds is $\sim 220$ km s$^{-1}$,
resembling the unorganized motion in the 
gravitational potential of isothermal 
mass distribution.

\begin{figure*}
\begin{center}
\includegraphics[width=17cm]{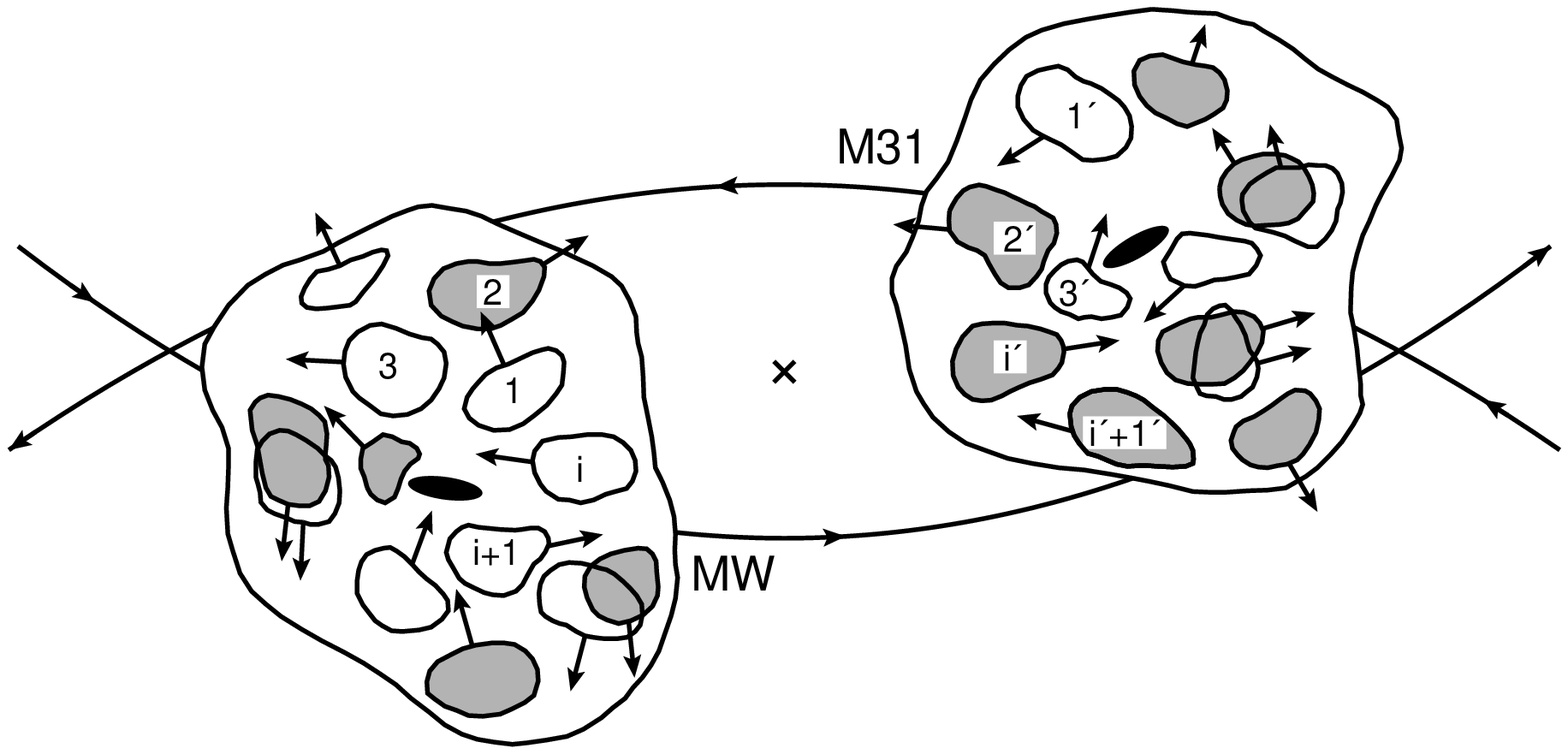}
\end{center}
\caption{Schematic view of model for 
primordial halos of the Galaxy and M31
at the time when they graze each other as 
in figure \ref{fig-5}(a). Baryonic gas 
fluctuations and dark matter fluctuations
are shown respectively by white and shaded 
contours, numbered with 
1, 2, $\cdots$, $i$, $i+1$, $\cdots$
in the Galaxy, and $1'$, $2'$, $\cdots$, $i'$, $i'+1'$,
$\cdots$ in M31. Some contours coincide 
spatially and others do not,
modeling the fact that the gravitational
force works similarly on the baryonic and
dark matters, but other forces
due to pressure, magnetic field and 
radiation work on the baryonic gas.
The solid lines indicate 
the orbits of the Galaxy and M31 about their
center of mass. The relative velocity is
$\sim 400$ km s$^{-1}$. Velocity arrows
attached to the clouds mimic the mean 
random velocity of approximately 220 km s$^{-1}$
in the proto-Galaxy halos and 250 km s$^{-1}$ in the proto-M31 halos.}
\label{fig-14}
\end{figure*}

From figures 14 and \ref{fig-5}(a), we can suppose
various types of collisions between these 
clouds. For example, the baryonic gas
clouds 1 and $1'$ would collide to
condense together leading to the formation 
of a small mass galaxy, or dwarf galaxy
with only a small amount of dark matter.
Another example is a high velocity collision 
of dark matter clouds, 2 and $2'$,
in the Galaxy halo
and the M31 halo, respectively. They would 
penetrate each other without generating a 
dark matter clump. When the colliding velocity 
is small like between $i$ and $i'$, or
between a baryonic and dark matter clouds,
the two-stream instability occurs to
form a gravitating matter clump or a
dwarf galaxy with a large amount of dark 
matter. The following dispersion relation given
in \citet{sas87} would be convenient for
these analyses,
\begin{equation}
\lambda _{\rm crit}={c^{2}-u^{2} \over
      \left[\langle v_{\rm d }^{2} 
      \rangle (\rho _{\rm g}/\rho _{\rm d})
       +2(c^{2}-u^{2})\right]^{1/2} }\lambda _{\rm Jd}. 
\end{equation}
Here $c$ is the sound velocity in the baryonic gas
which has the density of $\rho _{\rm g}$ and
streams with velocity of $u$ relative to the
dark matter whose density is $\rho _{\rm d}$
and velocity dispersion is $\langle v_{\rm d}^{2} \rangle$. 
The dark matter has a Jeans length given by 
$\lambda _{\rm Jd}$. When $u$ is large,
the collision would disperse the baryonic
and dark matter clumps, resulting in 
neither condensation nor formation of 
small mass galaxies.

We can assume other various collisions
such as multiple ones, $i$, $i+1$ and $i'$.
In other words, it is not so unrealistic 
that the dwarf member
of the LGG is a mixture of 
baryonic and dark matters in various
ratios. This dark matter model, shown 
schematically in figures 14 and \ref{fig-5}(a),
is, of course, {\it ad hoc} and far from conclusive, 
remaining very open to further 
investigations.

\section{Discussion and Conclusion} 

On the bases of the ring-like distributions of 
the Local Group of Galaxies and the Magellanic Stream
on the plane of sky (seen in the two-dimensional space), 
and of their coplanar distribution seen also 
in the three-dimensional 
space, we have proposed a dynamical model for the origin 
and evolution of 
the dwarf members of the LGG: The primordial dwarfs,
the primordial LMC and SMC as well, were driven to form 
in the high density region of gas 
generated hydrodynamically by the off-center collision between
the primordial gas-rich M31 and the primordial gas-rich
Galaxy some 10 Gyr ago. Newborn dwarf 
galaxies are scattered on the 
orbital plane of these two massive 
galaxies and they are trapped 
either to the potential well of the Galaxy or to that of M31, 
while other some number of dwarfs are left behind from such 
trappings. The latter objects are observed as if 
linking the Galaxy and M31 groups.

The orbital motion of M31 (relative to the Galaxy) is 
determined so that it successfully reproduces the 
well-studied dynamics and history of the LMC/SMC system. 
The most important model scenario is that the LMC/SMC are
born from the high-density gas of the formation site, 
together with other dwarf members of the LGG simultaneously. 
Thus the orbital plane of M31 must be parallel to that of the 
LMC/SMC, roughly normal to the line joining the 
present position of the sun and the Galactic center.
Both M31 and the LMC/SMC revolve round the Galaxy 
counterclockwise seen from the sun, on the unclosed 
elliptical orbit.

Thus we could have an answer to a long-standing 
question what the origin of the LMC/SMC system is and
where their large orbital angular momenta 
come from. Moreover interestingly, we could 
identify the birth places of M31 and the LMC/SMC 
on the sky and in the space, if precise age of 
the universe is given.

We are unable to demonstrate theoretically the validity of 
our dynamical model only from the data about the dwarf members:
their positions on the sky $(l, b)$, radial velocities and
radial distances measured from the Galactic center. 
Therefore, instead, 
we have searched for the dwarf members' past orbits that 
follow the model scenario. Figures \ref{fig-12} and \ref{fig-13} 
can actually show that the dwarf members start from their 
common formation site, move on the orbital plane (figure \ref{fig-13}),
and arrive at the present positions, reproducing the 
ring-like distribution in figures \ref{fig-3} and \ref{fig-4}.

From the results in Table \ref{tab-2} we find finally that more than 
sixty percent of the LGG members (22 out of 35) have $\bigcirc $,
suggesting our dynamical model for the LGG is very realistic.

The dwarf members marked with $\times$ in Table \ref{tab-2} do not
have their orbits that can be traced back in time to 
the formation site in our model scenario. 
They could be objects that plunged into the spherical LGG 
region from outside, or their orbits have been bent 
gravitationally by other nearby massive galaxies. 
We could also consider that these objects 
are intrinsic dwarfs born directly from the expanding 
universe and left behind from their merger to M31 
and/or to the Galaxy, although we have again a 
question why such cosmologically early-formed objects 
tend to gather in the LGG disk. 
  
On the other hand, if we relax our model conditions
in such a way that the LGG dwarfs are formed due also
to the tidal disturbance in gas around M31 and the
Galaxy, then the formation site depicted in 
figure \ref{fig-5}(a) would be largely extended on their
orbital plane. See a recent HST picture and its 
caption for dwarf formation in the Stepfan's
Quintet \citep{wea01}. The number
of $\bigcirc $ in Table \ref{tab-2} will increase, 
although it becomes difficult to predict 
the dwarfs' proper motions. 

A small mass spiral, M33, has been treated as a 
test-particle in determining the orbits of 
the LGG members. It is, of course, not correct in a sense 
that the mass of M33 is three to four times as large 
as that of the LMC/SMC system. It is necessary to 
join this spiral to our four-body problem,
but we do not yet have a key phenomenon and/or 
theory how to include M33 in a new 
five-body dynamics of 
the Galaxy-M31-M33-LMC-SMC in the LGG region. 
A fragmentary chain of hydrogen gas 
is observed as if bridging M33 and M31 
\citep{bli99}, and this may be usable to determine
the M33 orbit. However, we remain open to a conclusion 
if it is in the same category as the Magellanic 
Stream. Even if the gravity of M33 is taken into account,
however, we believe that our conclusions do not 
change so significantly, because the total
mass of the Galaxy and M31 dominates that of all
other members and control them gravitationally.

We did not use the data about the tidal-cut-off radii of
the LGG dwarfs, because, as figures \ref{fig-12} and \ref{fig-13} show, 
they could be located close to the Galaxy, or M31 or 
to both before they become condensed to dwarf galaxies.
Even if the pericenter distance of a dwarf's
orbit is calculated formally from the tidal-cut-off-radii
data, we cannot know what it means in our 
LGG model.

\citet{yoo02} found a group of metal-poor star-clusters
(counted as seven at the moment) and showed that
these clusters align on the plane perpendicular to the line joining the
present position of the sun and the Galactic center, in
quite a similar way to that discussed about the 
LGG members in figures \ref{fig-1} and \ref{fig-2}. As 
they suggest, if these metal-poor clusters were born originally 
in the LMC and have been captured recently
to the Galaxy, some of the globular clusters
in the Galactic halo are regarded as tightly 
related to the LGG dynamics that the LMC was driven to 
form at the off-center collision between M31 and the 
Galaxy.

We have pursued backward in time M31 and the LMC/SMC
in an expanding flat universe. When the present model
proves to be realistic and is improved to be more
quantitative, we must 
reexamine various parameters applied so far. 
For example, a claim may be made 
that the lifetime of 10.4 Gyr for the 
LGG model is a little too small and $\sim 12$ Gyr
more suitable. The former age has been obtained by use of 
the masses of the Galaxy and M31 which are applied 
uniformly in their series of papers
by \citet{mur80}, \citet{gar94}, 
\citet{fuj99} and \citet{saw99}. 
However, if we take
a-little-less masses for the Galaxy \citep{pra03} and M31, or their
smaller radii of dark matter halos, it is possible 
that the latter age is more reasonable. At present, 
however, we must wait for a moment to conclude 
which is more realistic and more
contribute to understanding the origin and 
evolution of the LGG.

We have only briefly touched upon some problems
such as the tidal effect on the disks of the Galaxy and M31, 
the dark matter mass-fraction among the LGG dwarfs, and etc. 
The extensive studies on these problems would 
contribute to obtaining new hints to improve and develop 
more realistically our LGG model.

\bigskip

We would like to thank Professor Y. Kumai for 
his critical comments and discussions 
about the present work, and one of us (M.F) deeply
thanks Vice-President of Nagoya University 
K. Yamashita for his hospitality and
research support during my stay at 
his U-Lab., Department of Physics and Astrophysics.

\end{document}